# An efficient framework for computing sensitivity of modal-related structural dynamic characteristics with multi-parameters

Kai Huang, Zhengguang Li*, Xiuli Wang

*School of Mathematics, Jilin University, Changchun, 130012, People's Republic of China*

**ABSTRACT**

The sensitivity of structural dynamic characteristics related to eigenmode (such as modal assurance criteria, modal flexibility, and modal mass etc.) has become a crucial and widely applied tool across various engineering fields. In this paper, a novel strategy is proposed for solving the sensitivity of structural dynamic characteristics related to eigenmode with respect to multiple variables. First, an algebraic method for computing the sensitivity of eigenvectors is developed to simplify the expression for sensitivity calculations. Subsequently, based on this new expression for eigenmode sensitivity, a framework for sensitivity analysis of structural dynamic characteristics related to eigenmodes with multiple parameters is established. With the incorporation of a preconditioning iterative method, the new computational framework effectively enhances the computational efficiency of sensitivity analysis for structural characteristics related to eigenmodes with multiple parameters. This framework is easy to operate and effectively reduces the "Fill-in" operations of sparse matrices. Three numerical examples are given to illustrate the effectiveness of the algorithm. The result shows that the novel strategy can significantly reduce central processing unit (CPU) computational time.

---

\* Corresponding author.
E-mail address: lizg@jlu.edu.cn (Zhengguang Li)

# 1. Introduction

Frequencies and mode shapes are critical indicators for characterizing the dynamic behavior of structures and play a pivotal role in understanding their response to dynamic loads. Structural dynamic characteristics related to eigenmode such as Modal Assurance Criterion( MAC) [1], Modal Strain Energy(MSE) [2] and Modal Flexibility (MF ) [3], Modal Mass[4] , Modal Stiffness[5] , Modal Participation Factor[6] , etc., are widely applied across various scientific and engineering fields, including model updating, structural optimization, dynamic modification, structural health monitoring and among others. The MAC is primarily used to measure the correlation between two modes and is currently employed in various applications. Gharehbaghi's review paper highlights that the MAC holds significant potential in recently developed model-based methodologies for the identification of mode shape data and the detection of changes in mode shapes of damaged structures[7]. Friswell et al.[8] considered using the MAC to compare the correlation between two modes. Ting et al.[9] used the MAC to correlate measurement data with numerical results and provided an example of correlating vibration test results of a Black Hawk helicopter with finite element model results. The MSE characterizes the strain energy of a structure caused by the modal vector. Cornwell et al.[10] proposed a method for locating damage in plate-like structures using MSE, which relies on the two-dimensional curvature of the structures and requires only the modal shapes before and after damage. Seyedpoor[11] effectively detected structural damage by utilizing MSE and computed the corresponding damage outcomes. Using MSE as a key damage indicator, Cha et al.[12] proposed a damage detection method for structural health monitoring in steel structures. Modal flexibility indicates that high-frequency modes contribute little to the flexibility matrix. As a result, the flexibility matrix can be efficiently approximated using only a limited number of low-frequency modes. This feature simplifies computations while retaining critical structural performance information, making it an efficient method in modal analysis[13]. Ni et al.[14] utilized modal flexibility to conduct health monitoring and damage localization for the cable-stayed Ting Kau Bridge. Jaishi et al. [15] used the modal flexibility method to identify damage in reinforced concrete beams in their research on structural health monitoring.

Sensitivity analysis is widely applied across various fields to evaluate how changes in input parameters affect system outputs. For structural applications, sensitivity derivations are not only important for design optimization, but also for system identification, statistical structural analysis and other related areas [16]. Mottershead et al. [17] demonstrated that sensitivity-based algorithms are particularly effective for tackling model updating problems. MAC sensitivity information is required by various commercial software programs, such as LMS Virtual.Lab and FEMtools, during model updating to ensure that the structural model more accurately reflects the actual structure[18]. Koh et al.[19] proposed the multiple damage location assurance criterion, which is sensitivity-based and can effectively reduce computational costs. Gomes et al.[20], in the field of damage identification, compared the modal sensitivity method with the genetic algorithm. Yan et al.[21] proposed a statistical structural damage detection method based on element modal strain energy sensitivity. Wu et al.[22] introduced a model updating approach utilizing the sensitivity of truncated modal flexibility, designed to address both system and local errors in the model.

The methods for solving sensitivity of structural dynamic characteristics related to eigenmode primarily involve two strategies: the forward mode and the adjoint mode. In the forward mode, the modal sensitivity is first calculated, followed by substituting it into the total derivative expression obtained via the chain rule, thereby deriving the sensitivity of the structural dynamic characteristics[23]. The forward mode has a simple computational format, but it is only suitable for situations with a small number of design variables. In cases involving multiple parameters, such as structural optimization, damage identification, model updating, etc., the computations in the forward mode can be prohibitively expensive. In the adjoint mode, the augmented function is first constructed, followed by the calculation of the sensitivities of structural dynamic characteristics after obtaining the adjoint variable[24-26]. This approach enables efficient analysis when multiple design variables are present. The forward mode and adjoint mode both involve techniques for computing modal sensitivities, even though modal sensitivities are not directly calculated in the adjoint mode. Many methods have been developed for calculating the modal sensitivity, including modal methods[27-30], Nelson's method[31-33], algebraic methods[34, 35], iterative method[36, 37], and so on. Li et al[38] introduced a novel approach for constructing a nonsingular coefficient matrix, which can be directly applied to compute the sensitivity of eigenvectors for both distinct and repeated eigenvalues.

This paper presents a novel strategy to address the sensitivity of structural dynamic characteristics related to eigenmodes with respect to multiple variables. An algebraic approach for calculating eigenvector derivatives is proposed to simplify sensitivity expressions. Based on this new eigenmode sensitivity expression, a sensitivity analysis framework for structural dynamic characteristics related to multi-parameter eigenmodes is established. This framework integrates a preconditioned iterative method while retaining the advantages of other mainstream frameworks. It effectively improves the computational efficiency of multi-parameter sensitivity analysis for eigenmode-related structural characteristics, simplifies operations, and reduces the "fill-in" effect in sparse matrices. Furthermore, this technique is suitable for integration into structural optimization and other CAE software.

The remainder of this paper is organized as follows. In section 2, the theoretical background of sensitivity of structural dynamic characteristics related to eigenmode with multi-parameters is introduced. Section 3 outlines the frameworks of the forward method, the adjoint method, and the newly proposed framework. Section 4 derives the sensitivities of three structural dynamic characteristics with respect to various variables. Section 5 provides three numerical examples to validate the effectiveness of the proposed algorithm. Finally, the paper concludes by summarizing the presented methodologies, highlighting their computational efficiency.

## 2. Theoretical background

For a freely undamped vibrating system with $N$ degrees of freedom (DOF), the generalized eigenvalue problems are typically expressed as

$$\mathbf{K}(\boldsymbol{p})\boldsymbol{\varphi}_i(\boldsymbol{p}) = \lambda_i(\boldsymbol{p})\mathbf{M}(\boldsymbol{p})\boldsymbol{\varphi}_i(\boldsymbol{p}), \qquad (i = 1,2,\cdots,N), \qquad (1)$$

where $\mathbf{M}(\boldsymbol{p}) \in \mathbb{R}^{N \times N}$ represents the mass matrix, $\mathbf{K}(\boldsymbol{p}) \in \mathbb{R}^{N \times N}$ represents the stiffness matrix, and $\mathbf{M}(\boldsymbol{p})$ is positive definite, $\mathbf{K}(\boldsymbol{p})$ is (semi-) positive definite. The elements of $\mathbf{M}(\boldsymbol{p})$ and $\mathbf{K}(\boldsymbol{p})$ are continuously dependent on design variable $\boldsymbol{p} \in \mathbb{R}^q$, and their elements $p_k$ $(k = 1,2 \cdots, q)$ may be material parameter, thicknesses and cross-sectional dimensions in model updating, or pseudo-density in structural topology optimization or stiffness reduction coefficient in structural health monitoring. $\lambda_i(\boldsymbol{p})$ is the $i$-th eigenvalue satisfying $\lambda_i = \omega_i^2$ the square of the $i$-th frequency in $rad/s$, $\boldsymbol{\varphi}_i(\boldsymbol{p})$ is the corresponding eigenvector, that can be normalized as follows,

$$\boldsymbol{\varphi}_i^{\mathrm{T}}(\boldsymbol{p})\mathbf{M}(\boldsymbol{p})\boldsymbol{\varphi}_j(\boldsymbol{p}) = \delta_{ij}, \qquad (i,j = 1,2,\cdots,N) \tag{2}$$

where $\delta_{ij}$ is the Kronecker delta.

Structural dynamic characteristics related to eigenmode $\mathcal{F} \in \mathbb{R}$ can be expressed as

$$\mathcal{F} = \mathcal{F}(\boldsymbol{p}, \lambda_i(\boldsymbol{p}), \boldsymbol{\varphi}_i(\boldsymbol{p})), \tag{3}$$

it is a function of the design variables $\boldsymbol{p}$, the eigenvalues $\lambda_i(\boldsymbol{p})$ and the mode shapes $\boldsymbol{\varphi}_i(\boldsymbol{p})$). In this function, changes in the design variables $p_k$ influence the stiffness and mass matrices, thereby altering the eigenmode and, subsequently, the structural dynamic characteristics $\mathcal{F}$. The sensitivity of dynamic characteristics can be obtained as follow by applying the chain rule,

$$\frac{d\mathcal{F}}{d\boldsymbol{p}} = \frac{\partial \mathcal{F}}{\partial \boldsymbol{p}} + \frac{\partial \mathcal{F}}{\partial \lambda_i}\frac{d\lambda_i}{d\boldsymbol{p}} + \left(\frac{\partial \mathcal{F}}{\partial \boldsymbol{\varphi}_i}\right)^{\mathrm{T}} \frac{d\boldsymbol{\varphi}_i}{d\boldsymbol{p}}. \tag{4}$$

where $d\mathcal{F}/d\boldsymbol{p} \in \mathbb{R}^q$ represents total derivative, and the $k$-th element of $d\mathcal{F}/d\boldsymbol{p}$ can be expressed as

$$\left(\frac{d\mathcal{F}}{d\boldsymbol{p}}\right)_k = \frac{\partial \mathcal{F}}{\partial p_k} + \frac{\partial \mathcal{F}}{\partial \lambda_i}\frac{\partial \lambda_i}{\partial p_k} + \left(\frac{\partial \mathcal{F}}{\partial \boldsymbol{\varphi}_i}\right)^{\mathrm{T}} \frac{\partial \boldsymbol{\varphi}_i}{\partial p_k}. \tag{5}$$

As shown in Eq. (5), calculating the sensitivity of structural dynamic characteristics involves computing the derivatives of both eigenvalues and eigenvectors. In practical engineering applications, the number of design variables $q$ tends to be large (for instance, in topology optimization, the number of design variables $q$ is typically comparable to the number of elements, reaching tens of thousands or even more). This situation results in a substantial computational workload. Many researchers have studied this issue and proposed various methods to effectively manage computational demands.

## 3. The framework for solving the sensitivity of structural dynamic characteristics related to eigenmode

### 3.1 The forward mode

In the forward mode, the sensitivities of eigenvalues (or frequencies) and mode shapes, $\partial \lambda_i / \partial p_k$ and $\partial \boldsymbol{\varphi}_i / \partial p_k$ are first calculated. These sensitivities are then substitutes into the total derivative expression in Eq. (3), thereby obtaining the sensitivities of structural dynamic characteristics.

Differentiating Eqs. (1) and (2) with respect to the design variable $p_k$, respectively, we yield

$$(\mathbf{K} - \lambda_i \mathbf{M}) \frac{\partial \boldsymbol{\varphi}_i}{\partial p_k} = -\left(\frac{\partial \mathbf{K}}{\partial p_k} - \frac{\partial \lambda_i}{\partial p_k} \mathbf{M} - \lambda_i \frac{\partial \mathbf{M}}{\partial p_k}\right) \boldsymbol{\varphi}_i, \tag{6}$$

and

$$\frac{\partial \boldsymbol{\varphi}_i^T}{\partial p_k} \mathbf{M} \boldsymbol{\varphi}_i + \boldsymbol{\varphi}_i^T \mathbf{M} \frac{\partial \boldsymbol{\varphi}_i}{\partial p_k} = -\boldsymbol{\varphi}_i^T \frac{\partial \mathbf{M}}{\partial p_k} \boldsymbol{\varphi}_i, \tag{7}$$

premultiplying Eq. (6) by $\boldsymbol{\varphi}_i^T$, and exploiting Eq. (2) we have

$$\frac{\partial \lambda_i}{\partial p_k} = \boldsymbol{\varphi}_i^T \left(\frac{\partial \mathbf{K}}{\partial p_k} - \lambda_i \frac{\partial \mathbf{M}}{\partial p_k}\right) \boldsymbol{\varphi}_i. \tag{8}$$

Note that the coefficient matrix $(\mathbf{K} - \lambda_i \mathbf{M})$ in Eq.6 is singular with the rank of $N - 1$, that cannot be inverted to solve the $\partial \boldsymbol{\varphi}_i / \partial p_k$.

    a.   **The forward mode with Nelson's Method**

The Nelson's method expresses the $\partial \boldsymbol{\varphi}_i / \partial p_k$ in the form of "homogeneous solution + particular solution":

$$\frac{\partial \boldsymbol{\varphi}_i}{\partial p_k} = \boldsymbol{\eta}_i + c_i \boldsymbol{\varphi}_i, \tag{9}$$

where $\boldsymbol{\eta}_i$ is the particular solution, $c_i$ is coefficient of the homogeneous solution. For convenience, denote $\mathbf{A}_i = \mathbf{K} - \lambda_i \mathbf{M}$ and $\mathbf{f}_i = -\left(\frac{\partial \mathbf{K}}{\partial p_k} - \frac{\partial \lambda_i}{\partial p_k}\mathbf{M} - \lambda_i \frac{\partial \mathbf{M}}{\partial p_k}\right)\boldsymbol{\varphi}_i$, $\boldsymbol{\eta}_i$ satisfies:

$$\mathbf{A}_i \boldsymbol{\eta}_i = \mathbf{f}_i. \tag{10}$$

The Nelson's method identifies the element with the largest absolute value in the eigenvector $\boldsymbol{\varphi}_i$, and denote it as $\varphi_{ji}$, where j represents the $j$-th component of $\boldsymbol{\varphi}_i$. Set all elements in the $j$-th row and $j$-th column of matrix $\mathbf{A}_i$ to zero, except for the diagonal element. Substitute the $j$-th diagonal element of $\mathbf{A}_i$ is set to the corresponding diagonal entry of the stiffness matrix $\mathbf{K}$, defining the updated matrix as $\bar{\mathbf{A}}_i$. Set the $j$-th component of $\mathbf{f}_i$, to zero, and refer to the resulting vector as $\bar{\mathbf{f}}_i$. The particular solutions can be expressed as

$$\boldsymbol{\eta}_i = (\bar{\mathbf{A}}_i)^{-1} \bar{\mathbf{f}}_i. \tag{11}$$

Further, substituting Eq. (11) to Eq. (7) yields

$$c_i = -\boldsymbol{\varphi}_i^T \mathbf{M} \boldsymbol{\eta}_i - \frac{1}{2}\boldsymbol{\varphi}_i^T \frac{\partial \mathbf{M}}{\partial p_k}\boldsymbol{\varphi}_i, \tag{12}$$

The sensitivity of the eigenvector is given by the following equation

$$\frac{\partial \boldsymbol{\varphi}_i}{\partial p_k} = (\bar{\mathbf{A}}_i)^{-1}\bar{\mathbf{f}}_i - \left(\boldsymbol{\varphi}_i^T \mathbf{M}\left((\bar{\mathbf{A}}_i)^{-1}\bar{\mathbf{f}}_i\right) + \frac{1}{2}\boldsymbol{\varphi}_i^T \frac{\partial \mathbf{M}}{\partial p_k}\boldsymbol{\varphi}_i\right)\boldsymbol{\varphi}_i, \tag{13}$$

Finally, the sensitivity of structural dynamic characteristics can be expressed as the following:

$$\left(\frac{d\mathcal{F}}{d\boldsymbol{p}}\right)_k = \frac{\partial \mathcal{F}}{\partial p_k} + \boldsymbol{\varphi}_i^T \left(\frac{\partial \mathbf{K}}{\partial p_k} - \lambda_i \frac{\partial \mathbf{M}}{\partial p_k}\right)\boldsymbol{\varphi}_i \frac{\partial \mathcal{F}}{\partial \lambda_i} + \left(\frac{\partial \mathcal{F}}{\partial \boldsymbol{\varphi}_i}\right)^T (\bar{\mathbf{A}}_i)^{-1}\bar{\mathbf{f}}_i$$
$$- \left(\frac{\partial \mathcal{F}}{\partial \boldsymbol{\varphi}_i}\right)^T \left(\boldsymbol{\varphi}_i^T \mathbf{M}\left((\bar{\mathbf{A}}_i)^{-1}\bar{\mathbf{f}}_i\right) + \frac{1}{2}\boldsymbol{\varphi}_i^T \frac{\partial \mathbf{M}}{\partial p_k}\boldsymbol{\varphi}_i\right)\boldsymbol{\varphi}_i \tag{14}$$

**b. The forward mode with algebraic method**

By combining Eq. (6) and Eq. (7), we can yield an algebraic equation for solving modal sensitivities

$$\begin{bmatrix} \mathbf{K} - \lambda_i \mathbf{M} & \mathbf{M}\boldsymbol{\varphi}_i \\ \boldsymbol{\varphi}_i^T \mathbf{M} & 0 \end{bmatrix} \begin{bmatrix} \frac{\partial \boldsymbol{\varphi}_i}{\partial p_k} \\ \frac{\partial \lambda_i}{\partial p_k} \end{bmatrix} = \begin{bmatrix} \frac{\partial \mathbf{K}}{\partial p_k}\boldsymbol{\varphi}_i - \frac{\partial \lambda_i}{\partial p_k}\mathbf{M}\boldsymbol{\varphi}_i \\ -\frac{1}{2}\boldsymbol{\varphi}_i^T \frac{\partial \mathbf{M}}{\partial p_k}\boldsymbol{\varphi}_i \end{bmatrix}. \tag{15}$$

Once $\partial\boldsymbol{\varphi}_i/\partial p_k$ and $\partial\lambda_i/\partial p_k$ are simultaneously determined by eq. (15), the final results can be derived using eq. (5). It can be observed that the fill-in operation on a large sparse matrix will significantly consume computational resources.

In addition, other methods for calculating modal sensitivities, such as modal method[39], iterative method[36] and others, can also be directly applied in the forward mode. The forward mode is conceptually straightforward and features a simple computational format, suitable for scenarios with a limited number of design variables. However, in cases involving multiple parameters, such as structural optimization, damage identification, and model updating, the computations in the forward mode can become prohibitively expensive, because it requires repeatedly solving system of linear equations (Eq.(10) or Eq.(15)).

**3.2 The framework of adjoint mode**

The adjoint method[24] fist define an augmented function

$$L = \mathcal{F} + \mathbf{v}^\mathrm{T}(\mathbf{K} - \lambda_i \mathbf{M})\boldsymbol{\varphi}_i + \alpha\left(\frac{1}{2}\boldsymbol{\varphi}_i^\mathrm{T}\mathbf{M}\boldsymbol{\varphi}_i - \frac{1}{2}\right), \tag{16}$$

where $\mathbf{v} \in \mathbb{R}^N$ and $\alpha \in \mathbb{R}$ are adjoint variables and need to be determined later. As $(\mathbf{K} - \lambda_i \mathbf{M})\boldsymbol{\varphi}_i = \mathbf{0}$ and $\boldsymbol{\varphi}_i^\mathrm{T}\mathbf{M}\boldsymbol{\varphi}_i - 1 = 0$ always holds true, differentiating the augmented function with respect to the design variable $p_k$, Eq. (17) can be obtained

$$\begin{aligned}\left(\frac{dL}{d\boldsymbol{p}}\right)_k &= \left(\frac{d\mathcal{F}}{d\boldsymbol{p}}\right)_k = \frac{\partial L}{\partial p_k} + \frac{\partial L}{\partial \lambda_i}\frac{\partial \lambda_i}{\partial p_k} + \left(\frac{\partial L}{\partial \boldsymbol{\varphi}_i}\right)^\mathrm{T}\frac{\partial \boldsymbol{\varphi}_i}{\partial p_k}\\
&= \frac{\partial \mathcal{F}}{\partial p_k} + \left(\frac{\partial \mathcal{F}}{\partial \lambda_i} - \mathbf{v}^\mathrm{T}\mathbf{M}\boldsymbol{\varphi}_i\right)\frac{\partial \lambda_i}{\partial p_k}\\
&\quad + \left(\mathbf{v}^\mathrm{T}(\mathbf{K} - \lambda_i\mathbf{M}) + \left(\frac{\partial \mathcal{F}}{\partial \boldsymbol{\varphi}_i}\right)^\mathrm{T} + \alpha\boldsymbol{\varphi}_i^\mathrm{T}\mathbf{M}\right)\frac{\partial \boldsymbol{\varphi}_i}{\partial p_k}\\
&\quad + \mathbf{v}^\mathrm{T}\left(\frac{\partial \mathbf{K}}{\partial p_k} - \lambda_i\frac{\partial \mathbf{M}}{\partial p_k}\right)\boldsymbol{\varphi}_i + \frac{1}{2}\alpha\boldsymbol{\varphi}_i^\mathrm{T}\frac{\partial \mathbf{M}}{\partial p_k}\boldsymbol{\varphi}_i,\end{aligned} \tag{17}$$

In Eq. (17), when the coefficients of $\frac{\partial \lambda_i}{\partial p_k}$ and $\frac{\partial \boldsymbol{\varphi}_i}{\partial p_k}$ are chosen to be zeros, which can be expressed as follows:

$$\frac{\partial L}{\partial \lambda_i} = \frac{\partial \mathcal{F}}{\partial \lambda_i} - \mathbf{v}^\mathrm{T}\mathbf{M}\boldsymbol{\varphi}_i = 0, \tag{18}$$

$$\frac{\partial L}{\partial \boldsymbol{\varphi}_i} = (\mathbf{K} - \lambda_i\mathbf{M})\mathbf{v} + \frac{\partial \mathcal{F}}{\partial \boldsymbol{\varphi}_i} + \alpha\mathbf{M}\boldsymbol{\varphi}_i = \mathbf{0}, \tag{19}$$

the differentiation of augmented functional can be solved:

$$\left(\frac{dL}{d\boldsymbol{p}}\right)_k = \frac{\partial \mathcal{F}}{\partial p_k} + \mathbf{v}^{\mathrm{T}}\left(\frac{\partial \mathbf{K}}{\partial p_k} - \lambda_i \frac{\partial \mathbf{M}}{\partial p_k}\right)\boldsymbol{\varphi}_i + \frac{1}{2}\alpha \boldsymbol{\varphi}_i^{\mathrm{T}} \frac{\partial \mathbf{M}}{\partial p_k}\boldsymbol{\varphi}_i, \qquad (20)$$

By premultiplying Eq. (19) by $\boldsymbol{\varphi}_i^{\mathrm{T}}$ and considering Eq. (2), adjoint variable $\alpha$ can be obtained as.

$$\alpha = -\boldsymbol{\varphi}_i^{\mathrm{T}} \frac{\partial \mathcal{F}}{\partial \boldsymbol{\varphi}_i}, \qquad (21)$$

Adjoint variables $\mathbf{v}$ satisfies the following equation:

$$(\mathbf{K} - \lambda_i \mathbf{M})\mathbf{v} = -\frac{\partial \mathcal{F}}{\partial \boldsymbol{\varphi}_i} - \alpha \mathbf{M}\boldsymbol{\varphi}_i, \qquad (22)$$

However due to the singularity of the coefficient matrix $(\mathbf{K} - \lambda_i \mathbf{M})$, the Eq. (22) cannot be solved directly. Nevertheless, the techniques for computing modal sensitivities can be employed to address this issue.

### a. The adjoint mode with Nelson's Method

Based on the traditional Nelson's method, the adjoint variable $\mathbf{v}$ in Eq. (22) be expressed as the sum of a particular solution and a homogeneous solution:

$$\mathbf{v} = \mathbf{v}_0 + c_i \boldsymbol{\varphi}_i, \qquad (23)$$

where $c_i$ are coefficients of the homogeneous solution and $\mathbf{v}_0$ is the particular solution, satisfies following equation:

$$(\mathbf{K} - \lambda_i \mathbf{M})\mathbf{v}_0 = -\left(\frac{\partial \mathcal{F}}{\partial \boldsymbol{\varphi}_i} + \alpha \mathbf{M}\boldsymbol{\varphi}_i\right), \qquad (24)$$

As mentioned earlier, Nelson's method forms a nonsingular coefficient matrix to compute $\mathbf{v}_0$, and the coefficients $c_i$ can be determined once the particular solution is obtained,

$$c_i = \frac{\partial \mathcal{F}}{\partial \lambda_i} - \boldsymbol{\varphi}_i^{\mathrm{T}} \mathbf{M}\mathbf{v}_0. \qquad (25)$$

Furthermore $\mathbf{v}$ and $d\mathcal{F}/dp_k$ can be sequentially determined using (23) and (20), respective. Algorithm 1 describes the calculation of the sensitivity of structural dynamic characteristics, related to eigenmodes, with respect to multiple variables using the adjoint mode with Nelson's method.

**Algorithm** 1 The adjoint mode with Nelson's method

1: $\alpha = -\boldsymbol{\varphi}_i^T \frac{\partial \mathcal{F}}{\partial \lambda_i}$.

2: Identify the element with the largest absolute value in the eigenvector $\boldsymbol{\varphi}_i$, and denote it as $\varphi_{ji}$, where $j$ represents the $j$-th component of $\boldsymbol{\varphi}_i$.

3: Denote $\mathbf{A}_i = \mathbf{K} - \lambda_i \mathbf{M}$, and set all elements in the $j$-th row and $j$-th column of $\mathbf{A}_i$ to zero, except for the diagonal element.

4: Replace the $j$-th diagonal element of $\mathbf{A}_i$ with the corresponding diagonal entry from the stiffness matrix $\mathbf{K}$, and denote the resulting updated matrix as $\bar{\mathbf{A}}_i$.

5: Set the $j$-th component of $\mathbf{f}_i = -(\partial \mathcal{F}/\partial \boldsymbol{\varphi}_i + \alpha \mathbf{M} \boldsymbol{\varphi}_i)$, to zero, and denote the resulting vector as $\bar{\mathbf{f}}_i$.

6: Solve $\mathbf{v}_0 = (\bar{\mathbf{A}}_i)^{-1} \bar{\mathbf{f}}_i$.

7: Solve $c_i = \frac{\partial \mathcal{F}}{\partial \lambda_i} - \boldsymbol{\varphi}_i^T \mathbf{M} \mathbf{v}_0$.

8: $\mathbf{v} = \mathbf{v}_0 + c_i \boldsymbol{\varphi}_i$.

9: For $k = 1, \cdots, q$, $q$ is the number of design parameters

10: $\left(\frac{d\mathcal{F}}{d\mathbf{p}}\right)_k = \frac{\partial \mathcal{F}}{\partial p_k} + \mathbf{v}^T \left(\frac{\partial \mathbf{K}}{\partial p_k} - \lambda_i \frac{\partial \mathbf{M}}{\partial p_k}\right) \boldsymbol{\varphi}_i + \frac{1}{2} \alpha \boldsymbol{\varphi}_i^T \frac{\partial \mathbf{M}}{\partial p_k} \boldsymbol{\varphi}_i$.

11: End

**b. The adjoint mode with algebraic method**

The algebraic method can be employed for find adjoint variables $\alpha$ and $\mathbf{v}$ in adjoint mode. Assembling Eq. (18) and (19), following algebraic equation can be constructed:

$$\begin{bmatrix} \mathbf{K} - \lambda_i \mathbf{M} & \mathbf{M} \boldsymbol{\varphi}_i \\ \boldsymbol{\varphi}_i^T \mathbf{M} & 0 \end{bmatrix} \begin{bmatrix} \mathbf{v} \\ \alpha \end{bmatrix} = \begin{bmatrix} -\frac{\partial \mathcal{F}}{\partial \boldsymbol{\varphi}_i} \\ \frac{\partial \mathcal{F}}{\partial \lambda_i} \end{bmatrix}, \quad (26)$$

The non-singularity of the coefficient matrix $\begin{bmatrix} \mathbf{K} - \lambda_i \mathbf{M} & \mathbf{M} \boldsymbol{\varphi}_i \\ \boldsymbol{\varphi}_i^T \mathbf{M} & 0 \end{bmatrix}$ can be observed if the eigenvalues $\lambda_i$ is distinct. Algorithm 2 describes the calculation of the sensitivity of structural dynamic characteristics, which are related to eigenmodes and multiple variables, using the adjoint mode with algebraic method.

**Algorithm 2** The adjoint mode with algebraic method

1: Construct the matrix $\begin{bmatrix} \mathbf{K} - \lambda_i \mathbf{M} & \mathbf{M} \boldsymbol{\varphi}_i \\ \boldsymbol{\varphi}_i^T \mathbf{M} & 0 \end{bmatrix}$.

2: Construct the right-hand side terms $\left[-\frac{\partial \mathcal{F}}{\partial \boldsymbol{\varphi}_i}, \frac{\partial \mathcal{F}}{\partial \lambda_i}\right]^T$.

3: Solve $\begin{bmatrix} \mathbf{K} - \lambda_i \mathbf{M} & \mathbf{M}\boldsymbol{\varphi}_i \\ \boldsymbol{\varphi}_i^T \mathbf{M} & 0 \end{bmatrix} \begin{bmatrix} \mathbf{v} \\ \alpha \end{bmatrix} = \begin{bmatrix} -\frac{\partial \mathcal{F}}{\partial \boldsymbol{\varphi}_i}, \frac{\partial \mathcal{F}}{\partial \lambda_i} \end{bmatrix}.$

4: For $k = 1, \cdots, q$, $q$ is the number of design parameters.

5: $\left(\frac{d\mathcal{F}}{d\mathbf{p}}\right)_k = \frac{\partial \mathcal{F}}{\partial p_k} + \mathbf{v}^T \left(\frac{\partial \mathbf{K}}{\partial p_k} - \lambda_i \frac{\partial \mathbf{M}}{\partial p_k}\right) \boldsymbol{\varphi}_i + \frac{1}{2}\alpha \boldsymbol{\varphi}_i^T \frac{\partial \mathbf{M}}{\partial p_k} \boldsymbol{\varphi}_i.$

6: End

Upon observation, the adjoint mode only requires solving a system of linear equations once (step 6 in **Algorithm** 1, step 2 in **Algorithm** 2), and it proves to be more efficient when the number of design variables exceed the number of response functions. However, both adjoint mode with Nelson's method and with algebraic method exhibit certain limitations when applied to large-scale complex problems. The adjoint mode with Nelson's method requires setting adjoint variables and expressing them as the sum of the particular solution and the homogeneous solution then solve each solution separately. The adjoint mod with algebraic method can directly solve for the adjoint variables once they are set. However, the system of algebraic equations generated by this method requires extensive 'fill-in' operations in the sparse matrix format, such as Compressed Sparse Row (CSR), which can be computationally expensive and memory-intensive for large and complex problems.

**3.3 New framework**

Form Eq. (7), Note that $(\partial \boldsymbol{\varphi}_i / \partial p_k)^T \mathbf{M} \boldsymbol{\varphi}_i$ is a scalar, we can derive:

$$\boldsymbol{\varphi}_i^T \mathbf{M} \frac{\partial \boldsymbol{\varphi}_i}{\partial p_k} = -0.5 \boldsymbol{\varphi}_i^T \frac{\partial \mathbf{M}}{\partial p_k} \boldsymbol{\varphi}_i, \tag{27}$$

For convenience, we denote $\mathbf{F} = \mathbf{K} - \lambda_i \mathbf{M}, \frac{\partial \mathbf{F}}{\partial p_k} = \frac{\partial \mathbf{K}}{\partial p_k} - \frac{\partial \lambda_i}{\partial p_k} \mathbf{M} - \lambda_i \frac{\partial \mathbf{M}}{\partial p_k}$, premultiplying Eq. (27) by $\mathbf{M}\boldsymbol{\varphi}_i$ and adding to Eq. (6), we obtain

$$\left(\mathbf{F} + \mathbf{M}\boldsymbol{\varphi}_i \boldsymbol{\varphi}_i^T \mathbf{M}\right) \frac{\partial \boldsymbol{\varphi}_i}{\partial p_k} = \left(-\frac{\partial \mathbf{F}}{\partial p_k} - 0.5 \mathbf{M}\boldsymbol{\varphi}_i \boldsymbol{\varphi}_i^T \frac{\partial \mathbf{M}}{\partial p_k}\right) \boldsymbol{\varphi}_i, \tag{28}$$

Denote $\left(\mathbf{F} + \mathbf{M}\boldsymbol{\varphi}_i \boldsymbol{\varphi}_i^T \mathbf{M}\right) = \mathbf{G}$, it will be proved that $\mathbf{G}$ is a non-singular matrix in the Appendix A, therefore we have

$$\frac{\partial \boldsymbol{\varphi}_i}{\partial p_k} = \mathbf{G}^{-1} \left(-\frac{\partial \mathbf{F}}{\partial p_k} - 0.5 \mathbf{M}\boldsymbol{\varphi}_i \boldsymbol{\varphi}_i^T \frac{\partial \mathbf{M}}{\partial p_k}\right) \boldsymbol{\varphi}_i \tag{29}$$

From the previous discussion, it is known that

$$\frac{\partial \lambda_i}{\partial p_k} = \boldsymbol{\varphi}_i^{\mathrm{T}} \left( \frac{\partial \mathbf{K}}{\partial p_k} - \lambda_i \frac{\partial \mathbf{M}}{\partial p_k} \right) \boldsymbol{\varphi}_i. \tag{30}$$

Therefore, Eq. (5) can be expressed as

$$\left( \frac{d\mathcal{F}}{d\boldsymbol{p}} \right)_k = \frac{\partial \mathcal{F}}{\partial p_k} + \boldsymbol{\varphi}_i^{\mathrm{T}} \left( \frac{\partial \mathbf{K}}{\partial p_k} - \lambda_i \frac{\partial \mathbf{M}}{\partial p_k} \right) \boldsymbol{\varphi}_i \frac{\partial \mathcal{F}}{\partial \lambda_i}$$

$$+ \left( \frac{\partial \mathcal{F}}{\partial \boldsymbol{\varphi}_i} \right)^{\mathrm{T}} \mathbf{G}^{-1} \left( -\frac{\partial \mathbf{F}}{\partial p_k} - 0.5 \mathbf{M} \boldsymbol{\varphi}_i \boldsymbol{\varphi}_i^{\mathrm{T}} \frac{\partial \mathbf{M}}{\partial p_k} \right) \boldsymbol{\varphi}_i. \tag{31}$$

For large and complex problems, $\mathbf{G}^{-1}$ is difficult to solve. Therefore, the operation of $\mathbf{G}^{-1}$ in the Eq. (31) is generally obtained by solving a linear system:

$$\mathbf{G}\mathbf{x} = \left( -\frac{\partial \mathbf{F}}{\partial p_k} - 0.5 \mathbf{M} \boldsymbol{\varphi}_i \boldsymbol{\varphi}_i^{\mathrm{T}} \frac{\partial \mathbf{M}}{\partial p_k} \right) \boldsymbol{\varphi}_i. \tag{32}$$

Note that the Eq. (32)'s right-hand side dependence on the design variables $p_k (k = 1, 2, \cdots, N)$, direct solving Eq. (31) by Eq. (32) is typically the most computationally expensive operation.

Denote $(\partial \mathcal{F} / \partial \boldsymbol{\varphi}_i)^{\mathrm{T}} \mathbf{G}^{-1} = \mathbf{y}^{\mathrm{T}}$, using symmetrical of $\mathbf{G}$ we can obtain

$$\mathbf{G}\mathbf{y} = \frac{\partial \mathcal{F}}{\partial \boldsymbol{\varphi}_i} \tag{33}$$

This linear system has no dependence on $p_k (k = 1, 2, \cdots, q)$, and the solution $\mathbf{y}$ is then used to compute $(d\mathcal{F} / d\boldsymbol{p})_k$

$$\left( \frac{d\mathcal{F}}{d\boldsymbol{p}} \right)_k = \frac{\partial \mathcal{F}}{\partial p_k} + \boldsymbol{\varphi}_i^{\mathrm{T}} \left( \frac{\partial \mathbf{K}}{\partial p_k} - \lambda_i \frac{\partial \mathbf{M}}{\partial p_k} \right) \boldsymbol{\varphi}_i \frac{\partial \mathcal{F}}{\partial \lambda_i}$$

$$+ \mathbf{y}^{\mathrm{T}} \left( -\frac{\partial \mathbf{F}}{\partial p_k} - 0.5 \mathbf{M} \boldsymbol{\varphi}_i \boldsymbol{\varphi}_i^{\mathrm{T}} \frac{\partial \mathbf{M}}{\partial p_k} \right) \boldsymbol{\varphi}_i. \tag{34}$$

Computing the sensitivity of structural dynamic characteristics by Eq. (34) requires only a single solution of the linear system (33). Consequently, this approach is particularly well-suited for sensitivity analysis involving multiple parameters.

The formation of the coefficient matrix $\mathbf{G}$ involves computing $\mathbf{M} \boldsymbol{\varphi}_i \boldsymbol{\varphi}_i^{\mathrm{T}} \mathbf{M}$, which introduces a significant number of fill-in operations. To address this issue, the SQMR algorithm [40] is employed with the preconditioner $\mathbf{K} - \mu \mathbf{M}$ to solve Eq. (33). As the factorization of $\mathbf{K} - \mu \mathbf{M}$ has been obtained during the process of solving eigenvalues and eigenvectors using the Lanczos algorithm or the subspace iteration method, with $\mu$ serves as a shift parameter, the preconditioner is constructed at no additional computational cost. The implementation of the SQMR algorithm for solving Eq. (33) is outlined below.

**Algorithm 3** SMQR algorithm

1: Choose $\mathbf{u}_0 \in \mathbb{R}^N$, and set the iteration error $\varepsilon$.
2: Compute $\mathbf{y} = \frac{\partial \mathcal{F}}{\partial \boldsymbol{\varphi}_i}$
3: Set $\mathbf{r}_0 = \mathbf{y} - \mathbf{G}\mathbf{u}_0$.
4: Solve $(\mathbf{K} - \mu\mathbf{M})\mathbf{t} = \mathbf{r}_0$.
5: Set $\tau_0 = \|\mathbf{t}\|_2, \mathbf{q}_0 = \mathbf{t}, v_0 = 0, d_0 = 0, and\ \rho_0 = \mathbf{r}_0^T\mathbf{q}_0$.
6: For $n = 1,2,\cdots$, do
7: $\mathbf{t} = \mathbf{G}\mathbf{q}_{n-1}$.
8: $\sigma_{n-1} = \mathbf{q}_{n-1}^T\mathbf{t}$, if $\sigma_{n-1} = 0$, then stop.
9: $\alpha_{n-1} = \frac{\rho_{n-1}}{\sigma_{n-1}}, \mathbf{r}_n = \mathbf{r}_{n-1} - \alpha_{n-1}\mathbf{t}$.
10: Solve $(\mathbf{K} - \mu\mathbf{M})\mathbf{t} = \mathbf{r}_n$
11: $v_n = \frac{\|\mathbf{t}\|_2}{\tau_{n-1}}, c_n = \frac{1}{\sqrt{1+v_n^2}}, \tau_n = \tau_{n-1}v_nc_n$.
12: $\mathbf{d}_n = c_n^2 v_{n-1}^2 \mathbf{d}_{n-1} + c_n^2 \alpha_{n-1}\mathbf{q}_{n-1}$.
13: $\mathbf{u}_n = \mathbf{u}_{n-1} + \mathbf{d}_n$
14: if $|\mathbf{u}_n - \mathbf{u}_{n-1}| < \varepsilon$, then break.
15: if $\rho_{n-1} = 0$, then break.
16: $\rho = \mathbf{r}_n^T\mathbf{t}, \beta_n = \frac{\rho_n}{\rho_{n-1}}, \mathbf{q}_n = \mathbf{t} + \beta_n\mathbf{q}_{n-1}$
17: End for

It should be noted that the coefficient matrix $\mathbf{K} - \mu\mathbf{M}$ in the above algorithm is factorized as $\mathbf{K} - \mu\mathbf{M} = \mathbf{L}\mathbf{D}\mathbf{L}^T$ during the eigenvalue and eigenvector computations. Therefore, the solution of $(\mathbf{K} - \mu\mathbf{M})\mathbf{t} = \mathbf{r}_0$ in this algorithm only requires forward and backward substitutions. Moreover, the matrix $\mathbf{G}$ does not need to be explicitly formed; instead, $\mathbf{G}\mathbf{q}_{n-1} = (\mathbf{F} + \mathbf{M}\boldsymbol{\varphi}_i\boldsymbol{\varphi}_i^T\mathbf{M})\mathbf{q}_{n-1}$ can be computed through matrix-vector multiplication.

The sensitivity of structural dynamic characteristics related to eigenmodes with respect to multiple variables, calculated using new framework is described in Algorithm 4.

**Algorithm 4** The method proposed in this paper

1: Solve $\mathbf{Gy} = \frac{\partial \mathcal{F}}{\partial \boldsymbol{\varphi}_i}$ Using Algorithm 3.

2: For $k = 1, \cdots, q$, $q$ is the number of design parameters

3: $\alpha = \mathbf{y}^T \left( -\frac{\partial \mathbf{F}}{\partial p_k} - 0.5 \mathbf{M} \boldsymbol{\varphi}_i \boldsymbol{\varphi}_i^T \frac{\partial \mathbf{M}}{\partial p_k} \right) \boldsymbol{\varphi}_i$.

4: $\beta = \boldsymbol{\varphi}_i^T \left( \frac{\partial \mathbf{K}}{\partial p_k} - \lambda_i \frac{\partial \mathbf{M}}{\partial p_k} \right) \boldsymbol{\varphi}_i$.

5: $\left( \frac{d\mathcal{F}}{d\mathbf{p}} \right)_k = \frac{\partial \mathcal{F}}{\partial p_k} + \beta \frac{\partial \mathcal{F}}{\partial \lambda_i} + \alpha$.

6: End for

## 4. Application to classic situations

Modal frequencies and mode shapes are fundamental characteristics of structural dynamic system, many structural dynamic characteristics related to eigenmode, such as the MAC, MSE, and MF, play a significant role in fields like aerospace, smart manufacturing, and maritime engineering. They are widely used in experimental modal analysis, structural model updating, health monitoring, and damage detection.

### 4.1 Modal assurance criterion

Modal assurance criterion is defined as follows

$$\mathcal{F}_{MAC} = \frac{\boldsymbol{\phi}_j^T \boldsymbol{\varphi}_i \boldsymbol{\varphi}_i^T \boldsymbol{\phi}_j}{(\boldsymbol{\phi}_j^T \boldsymbol{\phi}_j)(\boldsymbol{\varphi}_i^T \boldsymbol{\varphi}_i)}, \tag{35}$$

where $\boldsymbol{\phi}_j$, $\boldsymbol{\varphi}_i$ denotes the $j$-th modal vectors and $i$-th modal vectors. A MAC value closer to 1 means that the modes are more correlated, and a MAC value closer to 0 means that the modes are less correlated. Some partial derivatives involved in Eq. (31) can be expressed as the following:

$$\frac{\partial \mathcal{F}_{MAC}}{\partial p_k} = 0, \tag{36}$$

$$\frac{\partial \mathcal{F}_{MAC}}{\partial \lambda_i} = 0, \tag{37}$$

$$\frac{\partial \mathcal{F}_{MAC}}{\partial \boldsymbol{\varphi}_i} = 2 \left( \frac{(\boldsymbol{\varphi}_i^T \boldsymbol{\phi}_j) \boldsymbol{\phi}_j}{(\boldsymbol{\phi}_j^T \boldsymbol{\phi}_j)(\boldsymbol{\varphi}_i^T \boldsymbol{\varphi}_i)} - \frac{(\boldsymbol{\phi}_j^T \boldsymbol{\varphi}_i)(\boldsymbol{\varphi}_i^T \boldsymbol{\phi}_j) \boldsymbol{\varphi}_i}{(\boldsymbol{\phi}_j^T \boldsymbol{\phi}_j)(\boldsymbol{\varphi}_i^T \boldsymbol{\varphi}_i)^2} \right)^T. \tag{38}$$

### 4.2 Modal strain energy

The $r^{th}$ element modal strain energy is defined as

$$\mathcal{F}_{MSE_r} = \frac{1}{2}\boldsymbol{\varphi}_i^T \mathbf{K}_r \boldsymbol{\varphi}_i, \tag{39}$$

where $\mathbf{K}_r$ is the $r^{th}$ element stifness matrix, $\boldsymbol{\varphi}_i$ denotes the $i$-th modal vector. Some partial derivatives involved in Eq. (31) can be expressed as the following:

$$\frac{\partial \mathcal{F}_{MSE_r}}{\partial p_k} = \frac{1}{2}\boldsymbol{\varphi}_i^T \frac{\partial \mathbf{K}_r}{\partial p_k} \boldsymbol{\varphi}_i, \tag{40}$$

$$\frac{\partial \mathcal{F}_{MSE_r}}{\partial \lambda_i} = 0, \tag{41}$$

$$\frac{\partial \mathcal{F}_{MSE_r}}{\partial \boldsymbol{\varphi}_i} = \mathbf{K}_r \boldsymbol{\varphi}_i. \tag{42}$$

**4.3 Modal flexibility**

The definition of modal flexibility is

$$\mathcal{F}_{MF} = \boldsymbol{\varphi}_i^T \frac{1}{\omega_i^2} \boldsymbol{\varphi}_i = \boldsymbol{\varphi}_i^T \frac{1}{\lambda_i} \boldsymbol{\varphi}_i, \tag{43}$$

where $\boldsymbol{\varphi}_i$ denotes the $i$-th modal vector, $\omega_i$ denotes the $i$-th order frequency. Similar to the approach in the previous sections, some partial derivatives involved in Eq. (31) can be expressed as the following:

$$\frac{\partial \mathcal{F}_{MF}}{\partial p_k} = 0, \tag{44}$$

$$\frac{\partial \mathcal{F}_{MF}}{\partial \lambda_i} = -\frac{1}{\lambda_i^2} \boldsymbol{\varphi}_i^T \boldsymbol{\varphi}_i, \tag{45}$$

$$\frac{\partial \mathcal{F}_{MF}}{\partial \boldsymbol{\varphi}_i} = 2\frac{1}{\lambda_i} \boldsymbol{\varphi}_i. \tag{46}$$

**5. Numerical example**

In this section, three examples are given to illustrate the efficiency of the proposed method. The Intel® Math Kernel Library 2024.2 was used as the underlying math library, with an Intel® Core™ i7-14700KF processor and 64GB of memory. The pseudo-density $\rho_i$, often employed in the fields of topology optimization and damage identification, is chosen as a design parameter. The initial value of the pseudo-density is set to 1, and the corresponding global stiffness matrix and mass matrix are shown below.

$$\mathbf{K} = \sum_{i=1}^{q} \rho_i^3 \mathbf{K}_{ei}, \mathbf{M} = \sum_{i=1}^{q} \rho_i \mathbf{M}_{ei}, \tag{47}$$

where, **K** is the stiffness matrix, **M** is the mass matrix, $\mathbf{K}_{ei}$ and $\mathbf{M}_{ei}$ denote the element stiffness matrix and element mass matrix, $q$ is the total number of elements, respectively. In the following numerical examples, all cases use the proposed method with consistent parameters. For the SMQR algorithm, the iterative tolerance is set to $10^{-5}$.

The relative error of the sensitivity is defined as follows:

$$e_r(\%) = \left(\frac{|s_p - s_n|}{|s_n|}\right) \times 100\%, \qquad (48)$$

where $s_p$ and $s_n$ denote sensitivity calculated by using present method and the traditional method, respectively. The advantage of the present method over another is measured by the CPU time, the efficiency ratio is considered and defined as follows:

$$R_{AB} = \frac{T_A}{T_B} \qquad (49)$$

where $T_A$ represents the CPU time of algorithm A, and $T_B$ represents the CPU time of algorithm B.

**Example 1**

The plane vibration of rectangular plate structure with all four corners clamped is considered, as shown in Fig. 1. The lengths of plate sides are X and Y respectively. We continuously expand the geometric scale of the plate and perform mesh discretization. The Young's modulus of material is $E = 2 \times 10^{11} Pa$, the Poisson's ratio is $\gamma = 0.3$, and the densities is $\rho = 7.8 \times 10^3 Kg/m^3$.

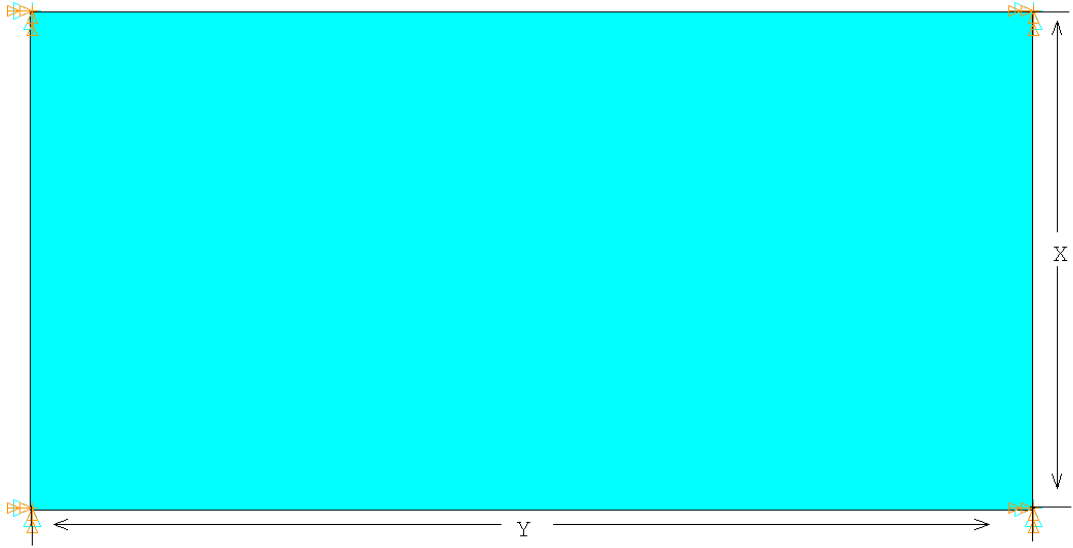

**FIGURE 1** The plate structural

The sensitivity of structural dynamic characteristics, MAC, MSE, and MF are solved by using the present method (PM), the adjoint mode with algebraic method (ADAM), the Nelson method (NE), and the adjoint mode with Nelson's method (ADNE). The CPU time for sensitivity calculations using different method was measured by averaging the results over twenty repeated computations.

The comparison of CPU times for the four methods under different geometric dimensions are listed in Tables 1, 3, and 5, the $L^{\infty}$ norm of sensitivities of the structural dynamic characteristics with respect to each design variable are shown in Tables 2, 4, and 6. Since the Nelson method provides exact results, we compare our method with the Nelson method in terms of relative error (Eq.48).

**TABLE 1** Comparison of CPU times for calculating MAC sensitivity

| X($m$) | Y($m$) | DOFs | ADAM(s) | NE(s) | ADNE(s) | PM(s) |
|---|---|---|---|---|---|---|
| 20 | 10 | 462 | 0.001795 | 0.008057 | 0.001199 | 0.000359 |
| 40 | 10 | 902 | 0.002978 | 0.027363 | 0.001924 | 0.000543 |
| 40 | 30 | 2542 | 0.005913 | 0.314869 | 0.004150 | 0.002365 |

| 60 | 50 | 6222 | 0.011960 | 1.369373 | 0.010889 | 0.004458 |
| 80 | 70 | 11502 | 0.024369 | 3.258172 | 0.020208 | 0.008084 |
| 100 | 80 | 16362 | 0.041049 | 5.095644 | 0.028499 | 0.009802 |
| 120 | 100 | 24442 | 0.056749 | 6.718391 | 0.036965 | 0.014012 |
| 140 | 120 | 34122 | 0.085271 | 11.157651 | 0.054545 | 0.016521 |
| 180 | 140 | 51042 | 0.125712 | 38.162817 | 0.087641 | 0.023111 |

**TABLE 2** Accuracy comparison of different methods for calculating MAC sensitivity

| X($m$) | Y($m$) | DOFs | ADAM | NE | ADNE | PM | RE (%) |
|---|---|---|---|---|---|---|---|
| 20 | 10 | 462 | -0.843790 | -0.843790 | -0.843790 | -0.843791 | 0.0001 |
| 40 | 10 | 902 | -0.208972 | -0.208972 | -0.208972 | -0.209000 | 0.0135 |
| 40 | 30 | 2542 | -2.889908 | -2.889908 | -2.889908 | -2.889916 | 0.0003 |
| 60 | 50 | 6222 | -4.954368 | -4.954368 | -4.954368 | -4.954379 | 0.0002 |
| 80 | 70 | 11502 | -7.029071 | -7.029071 | -7.029070 | -7.029081 | 0.0001 |
| 100 | 80 | 16362 | -4.153120 | -4.153120 | -4.153120 | -4.153127 | 0.0002 |
| 120 | 100 | 24442 | -5.230679 | -5.230679 | -5.230678 | -5.230685 | 0.0001 |
| 140 | 120 | 34122 | -6.306619 | -6.306619 | -6.306618 | -6.306624 | 0.0001 |
| 180 | 140 | 51042 | -3.777403 | -3.777403 | -3.777402 | -3.777408 | 0.0001 |

**TABLE 3** Comparison of CPU times for calculating MSE sensitivity

| X($m$) | Y($m$) | DOFs | ADAM(s) | NE(s) | ADNE(s) | PM(s) |
|---|---|---|---|---|---|---|
| 20 | 10 | 462 | 0.001714 | 0.008595 | 0.001230 | 0.000464 |
| 40 | 10 | 902 | 0.003406 | 0.028099 | 0.002842 | 0.000736 |
| 40 | 30 | 2542 | 0.005857 | 0.309514 | 0.004640 | 0.003043 |
| 60 | 50 | 6222 | 0.012116 | 1.375109 | 0.011081 | 0.005222 |
| 80 | 70 | 11502 | 0.023473 | 3.317845 | 0.021824 | 0.009833 |
| 100 | 80 | 16362 | 0.042113 | 5.003873 | 0.030252 | 0.011931 |
| 120 | 100 | 24442 | 0.056080 | 6.780057 | 0.038593 | 0.017425 |
| 140 | 120 | 34122 | 0.085375 | 10.566350 | 0.057940 | 0.020889 |
| 180 | 140 | 51042 | 0.125301 | 36.444274 | 0.086848 | 0.024169 |

**TABLE 4** Accuracy comparison of different methods for calculating MSE sensitivity

| X($m$) | Y($m$) | DOFs | ADAM | NE | ADNE | PM | RE(%) |
|---|---|---|---|---|---|---|---|
| 20 | 10 | 462 | -987.93209 | -987.93209 | -987.93209 | -987.93205 | 0.000003 |
| 40 | 10 | 902 | -95.69792 | -95.69792 | -95.69792 | -95.69790 | 0.000022 |
| 40 | 30 | 2542 | -274.07663 | -274.07663 | -274.07663 | -274.07663 | 0.000001 |
| 60 | 50 | 6222 | -136.11771 | -136.11771 | -136.11771 | -136.11818 | 0.000345 |
| 80 | 70 | 11502 | -85.82304 | -85.82304 | -85.82304 | -85.82337 | 0.000390 |
| 100 | 80 | 16362 | -37.44550 | -37.44550 | -37.44550 | -37.44556 | 0.000148 |
| 120 | 100 | 24442 | -27.69997 | -27.69997 | -27.69997 | -27.70005 | 0.000267 |
| 140 | 120 | 34122 | -21.65892 | -21.65892 | -21.65892 | -21.65890 | 0.000106 |
| 180 | 140 | 51042 | -9.20372 | -9.20372 | -9.20372 | -9.20057 | 0.034259 |

**TABLE 5** Comparison of CPU times for calculating MF sensitivity

| X(m) | Y(m) | DOFs | ADAM(s) | NE(s) | ADNE(s) | PM(s) |
|---|---|---|---|---|---|---|
| 20 | 10 | 462 | 0.001485 | 0.008663 | 0.001306 | 0.000350 |
| 40 | 10 | 902 | 0.003393 | 0.028558 | 0.002580 | 0.000580 |
| 40 | 30 | 2542 | 0.005640 | 0.305208 | 0.004569 | 0.001954 |
| 60 | 50 | 6222 | 0.012061 | 1.369619 | 0.013797 | 0.003562 |
| 80 | 70 | 11502 | 0.023334 | 3.293164 | 0.022018 | 0.006235 |
| 100 | 80 | 16362 | 0.042431 | 5.184623 | 0.030922 | 0.008060 |
| 120 | 100 | 24442 | 0.056662 | 6.766742 | 0.038315 | 0.010005 |
| 140 | 120 | 34122 | 0.089427 | 10.878269 | 0.060441 | 0.014227 |
| 180 | 140 | 51042 | 0.127649 | 35.571024 | 0.087747 | 0.019138 |

**TABLE 6** Accuracy comparison of different methods for calculating MF sensitivity

| X(m) | Y(m) | DOFs | ADAM | NE | ADNE | PM | RE (%) |
|---|---|---|---|---|---|---|---|
| 20 | 10 | 462 | -4.7198E-10 | -4.7198E-10 | -4.7198E-10 | -4.7199E-10 | 0.0012 |
| 40 | 10 | 902 | -1.6425E-09 | -1.6425E-09 | -1.6425E-09 | -1.6425E-09 | 0.0008 |
| 40 | 30 | 2542 | -2.5373E-09 | -2.5373E-09 | -2.5373E-09 | -2.5373E-09 | 0.0005 |
| 60 | 50 | 6222 | -6.2345E-09 | -6.2345E-09 | -6.2345E-09 | -6.2345E-09 | 0.0002 |
| 80 | 70 | 11502 | -1.1557E-08 | -1.1557E-08 | -1.1557E-08 | -1.1556E-08 | 0.0079 |
| 100 | 80 | 16362 | -1.6521E-08 | -1.6521E-08 | -1.6521E-08 | -1.6520E-08 | 0.0037 |
| 120 | 100 | 24442 | -2.4694E-08 | -2.4694E-08 | -2.4694E-08 | -2.4694E-08 | 0.0019 |
| 140 | 120 | 34122 | -3.4497E-08 | -3.4497E-08 | -3.4497E-08 | -3.4497E-08 | 0.0011 |
| 180 | 140 | 51042 | -5.1853E-08 | -5.1853E-08 | -5.1853E-08 | -5.1852E-08 | 0.0005 |

It is found that the relative errors are very small, with the maximum relative error being only 0.034%. The CPU time of the present method is considerably efficient with respect to forward mode and the adjoint mode with Nelson' and algebraic method. As the degrees of freedom increase, the efficiency difference will become even more pronounced. Figure 2 shows that the method precented in this paper achieves increasing speed advantages as the scale expands.

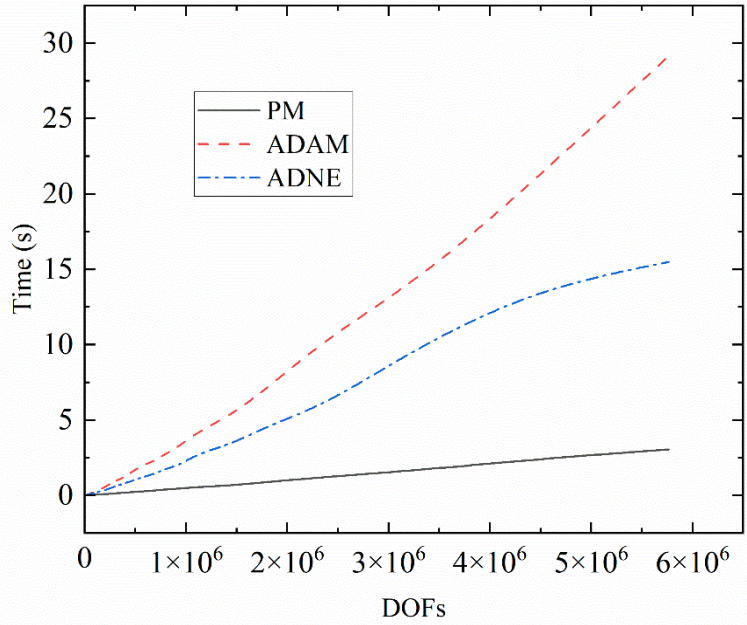

(a) Comparison of Three Different Methods for Calculating MAC Function Sensitivity

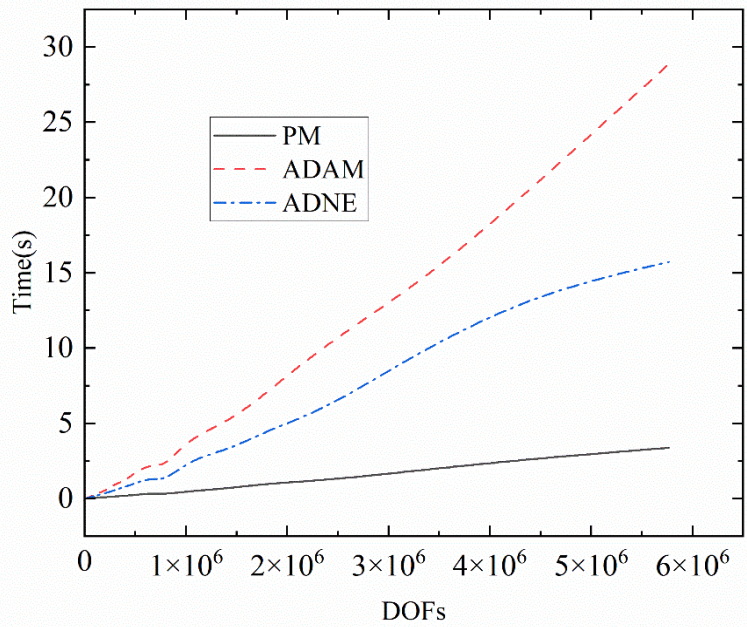

(b) Comparison of Three Different Methods for Calculating MF Function Sensitivity

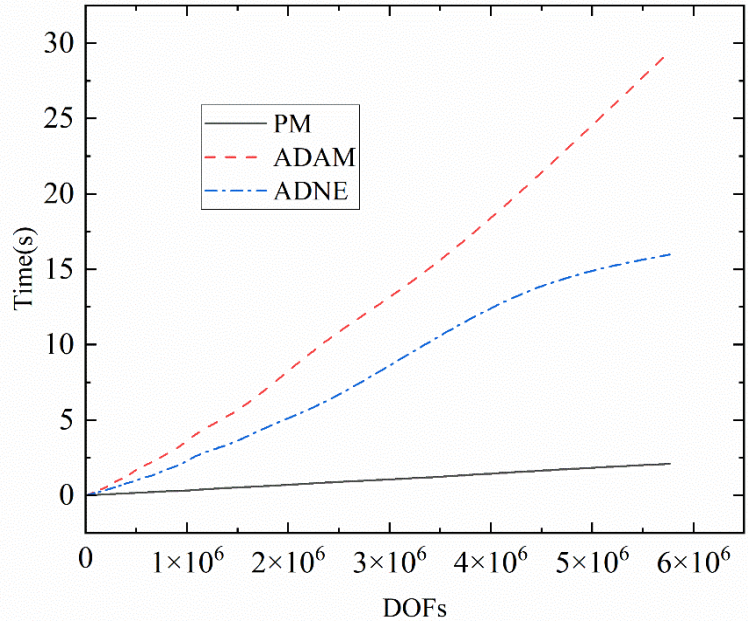

(c) Comparison of Three Different Methods for Calculating MSE Function Sensitivity

**FIGURE 2** Comparison of Sensitivity Calculation CPU Times for Different Performance Metrics

**Example 2**

A finite element model of aircraft wing structure, as shown Figure 3. This model consists of two main components: the wing framework and the skin. The material properties of wing framework are Young's Modulus $E = 1.2 \times 10^{12}$ $Pa$, Poisson's Ratio $\gamma = 0.3$, density $\rho = 4500 kg/m^3$, the wing skin material are Young's Modulus $E = 2.6 \times 10^8$ $Pa$, Poisson's Ratio is $\gamma = 0.3$, density $\rho = 886 kg/m^3$. The wing framework is modeled using 3D beam elements, while the aircraft skin is represented by shell elements. The finite element model consists of 3,385 beam elements, 39,739 shell elements and 39,721 nodes. The total number of degrees of freedom are 238,326.

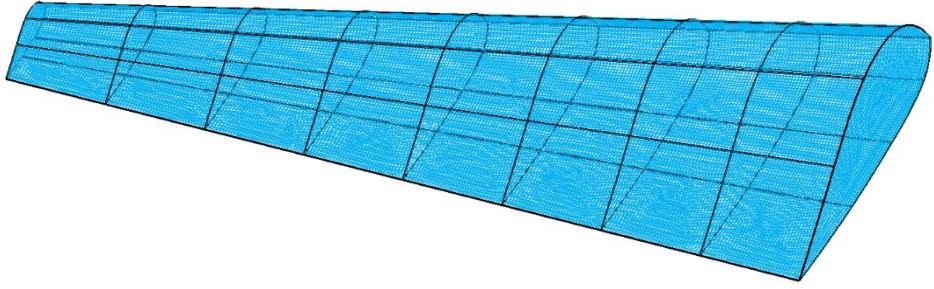

**FIGURE 3** The finite element mesh of the aircraft wing

In this case the sensitivity of modal flexibility of the first-order modal for the aircraft wing is calculated by using the present method (PM), the adjoint mode with algebraic method (ADAM), and the adjoint mode with Nelson's method (ADNE). Comparison of CPU times of the PM, ADAM and ADNE is shown in Table 7. It is found that, compared with CPU run times of the ADAM and ADNE, the efficiency ratio of the present method is above 2.47, the present algorithm significantly reduces computation time.

**TABLE 7** Comparison of CPU times

| Method | CPU time(s) | $R_{AB}$ |
|--------|-------------|----------|
| PM     | 0.5505      |          |
| ADAM   | 2.0515      | 3.7267   |
| ADNE   | 1.3643      | 2.4783   |

Select the 50th element here, and list the MSE sensitivities of the preceding 5 elements and the following 5 elements in Table 8. Using the results of ADAM and ADNE as a reference, it can be observed that the relative errors are controlled within $1.0 \times 10^{-7}\%$. Comparisons of the contour maps for the sensitivity of the first-order modal flexibility is presented in Figure 4. It is found that the results of three methods are also very close.

**TABLE 8** Accuracy Comparison

| Element Number | ADAM | AMNE | PM | Relative Error (%) |
|---|---|---|---|---|
| 45 | 0.02662765034 | 0.02662765034 | 0.02662765035 | 4.80831E-08 |
| 46 | 0.03194068370 | 0.03194068370 | 0.03194068371 | 4.10705E-08 |
| 47 | 0.03890936455 | 0.03890936455 | 0.03890936457 | 3.70474E-08 |
| 48 | 0.04775726836 | 0.04775726836 | 0.04775726838 | 3.39483E-08 |
| 49 | 0.05879101461 | 0.05879101461 | 0.05879101463 | 3.29486E-08 |
| 50 | -1.91584707710 | -1.91584707710 | -1.91584707707 | 1.3571E-09 |
| 51 | 0.08857633471 | 0.08857633471 | 0.08857633474 | 3.29484E-08 |
| 52 | 0.10809418468 | 0.10809418468 | 0.10809418472 | 3.33977E-08 |
| 53 | 0.13128175276 | 0.13128175276 | 0.13128175280 | 3.40984E-08 |
| 54 | 0.15850789430 | 0.15850789430 | 0.15850789435 | 3.49257E-08 |
| 55 | 0.19010914191 | 0.19010914191 | 0.19010914198 | 3.5588E-08 |

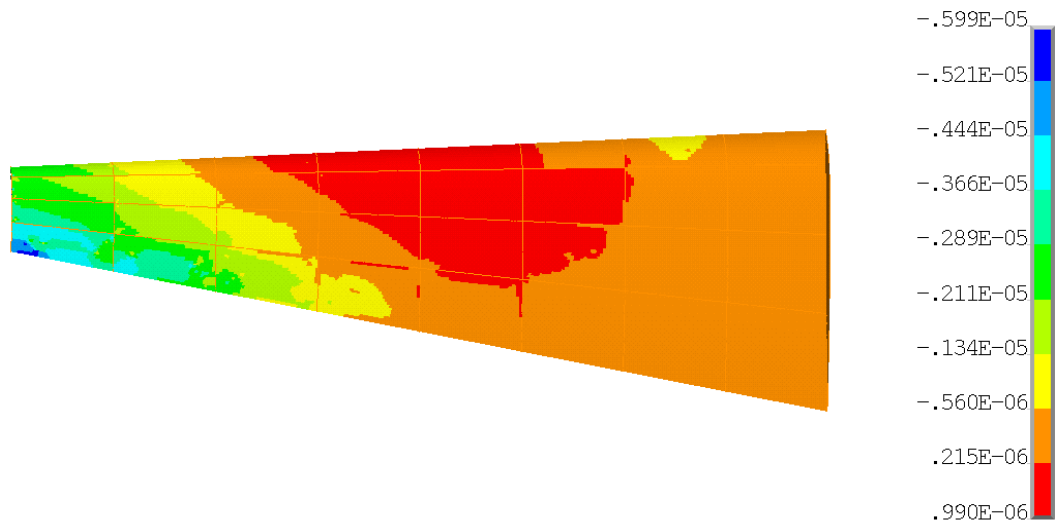

(a) The results of calculating modal strain energy sensitivity using PM

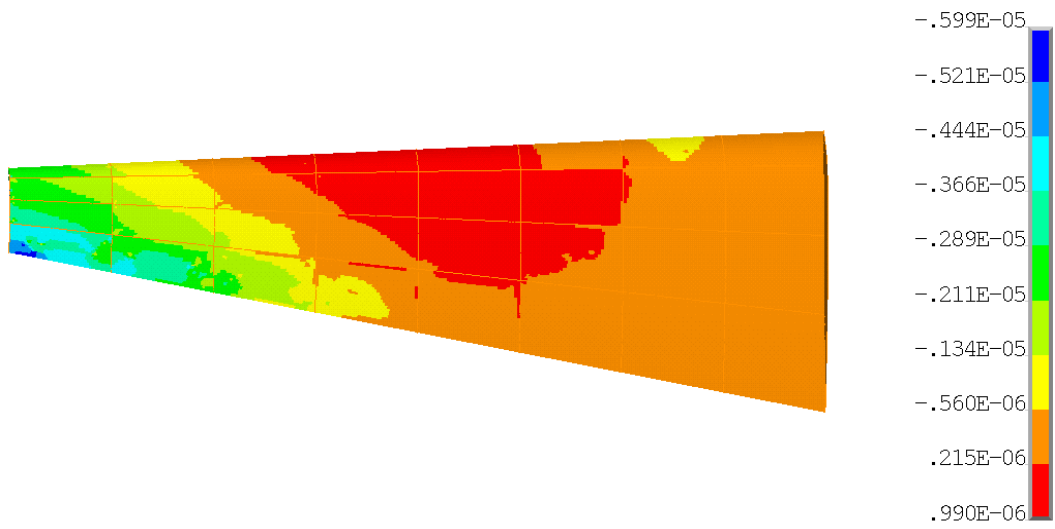

(b) The results of calculating modal strain energy sensitivity using ADAM

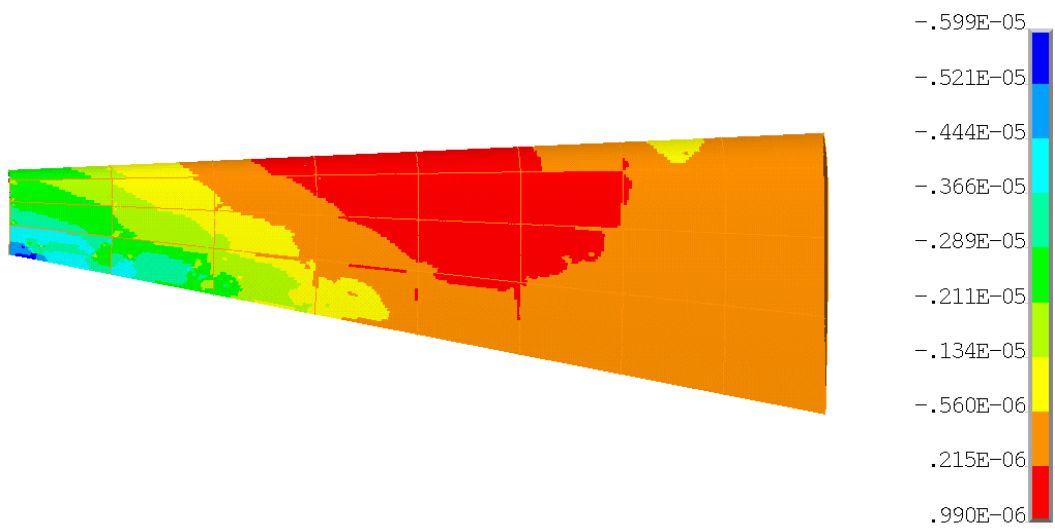

(c) The results of calculating modal strain energy sensitivity using ADNE

**FIGURE 4** The results of calculating modal flexibility sensitivity using the three different methods

**Example 3**

A Body in White (BIW) model of a vehicle, as shown in the Figure 5, its' exploded view of BIW show as Figure 6. The Young's modulus of material is $E = 2 \times 10^{11} \text{Pa}$, the Poisson's ratio is $\gamma = 0.3$, and the densities is $\rho = 7.8 \times 10^3 \text{Kg}/m^3$. A finite element model is established that consists of 129,983 nodes and 126,162 three-dimensional shell elements. The total number of degrees of freedom are 779,898.

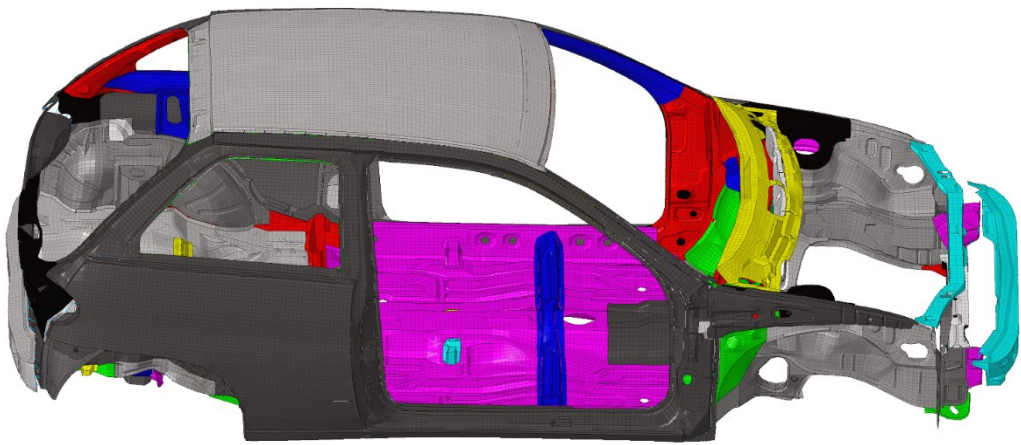

**FIGURE 5** Body in White model

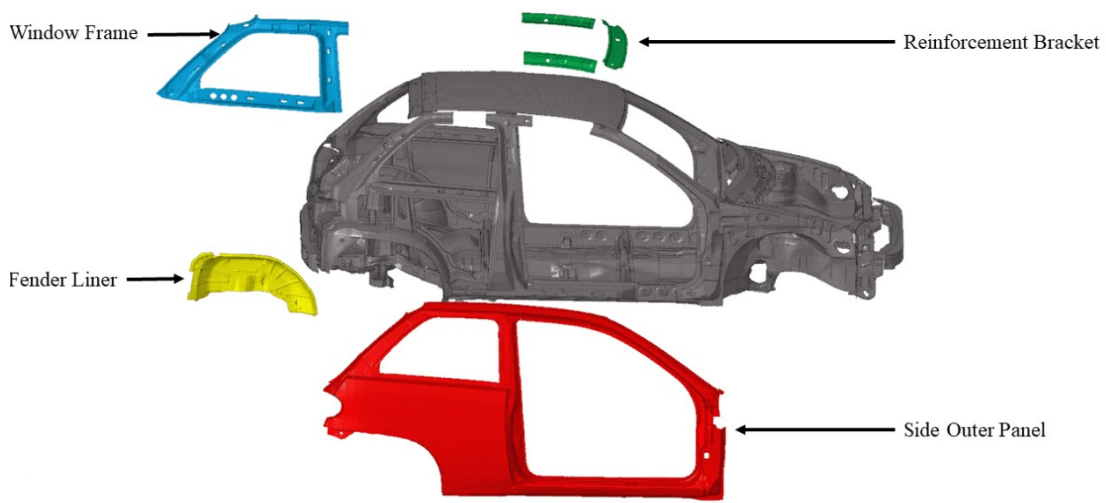

**FIGURE 6** Exploded View of the Body in White

In this case the sensitivity of modal flexibility of the first-order modal for the BIW is calculated by using the PM, the ADAM and the ADNE. Comparison of CPU run times of the PM, ADAM and ADNE is shown in Table 9. It is found that, compared with CPU run times of the ADAM and ADNE, the efficiency ratio of the present method is above 3.04, the present algorithm significantly reduces computation time.

**TABLE 9** Efficiency Comparison

| Method | Time(s) | $R_{AB}$ |
|---|---|---|
| PM | 1.5220 | |
| ADAM | 7.8045 | 5.1276 |
| ADNE | 4.6291 | 3.0414 |

As in the previous example, the $L^\infty$ norm of sensitivity of modal flexibility with respect to each design variable are listed in Table 10. Using the results of ADAM and ADNE as a reference, it can be observed that the relative errors are very small with $2.434 \times 10^{-3}\%$.

**TABLE 10** Efficiency Comparison

| Method | Sensitivity of modal flexibility | Relative Error (%) |
|---|---|---|
| PM | −0.7117200 | $2.434 \times 10^{-3}$ |
| ADAM | −0.7117027 | |
| ADNE | −0.7117027 | |

The contour maps for the sensitivity of the first-order modal flexibility by using PM is presented in Figure 7. For clarity and better comparison, four components with significant influence, namely the Side Outer Panel, Window Frame, Fender Liner, and Reinforcement Bracket are selected as show in Figure 8 and Figure 9. It is found that the results of three methods are also very close.

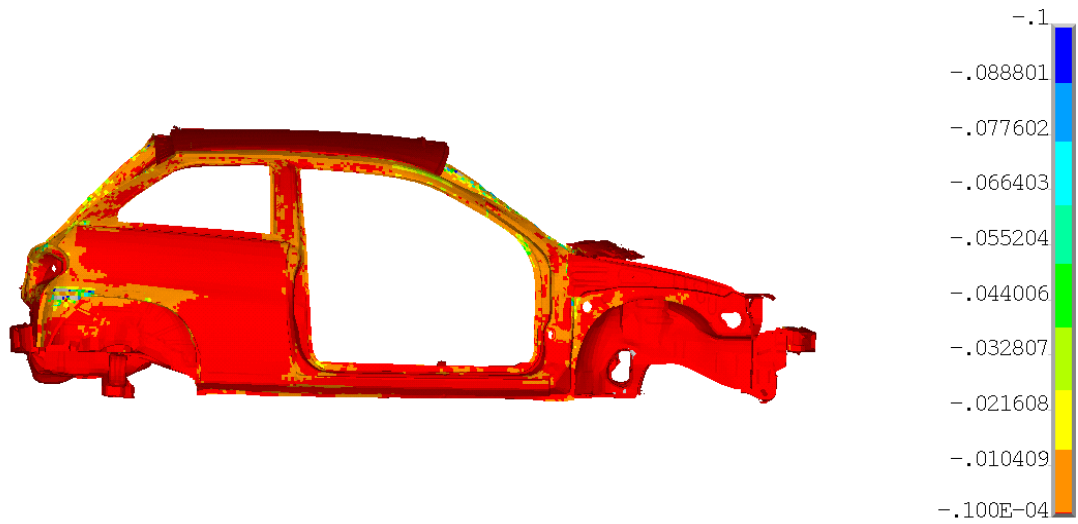

**FIGURE 7** The global result visualization

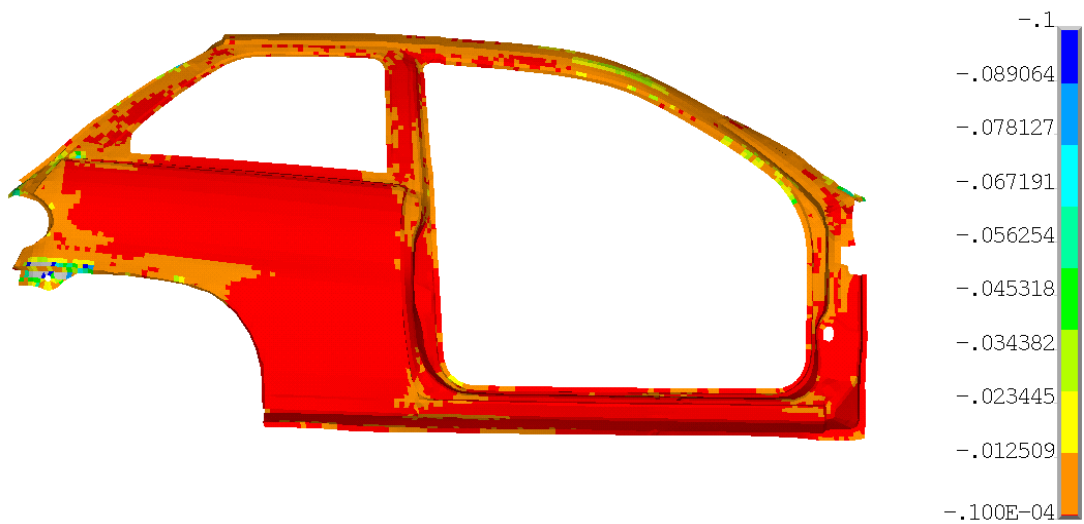

(a) The results of calculating modal flexibility sensitivity using PM

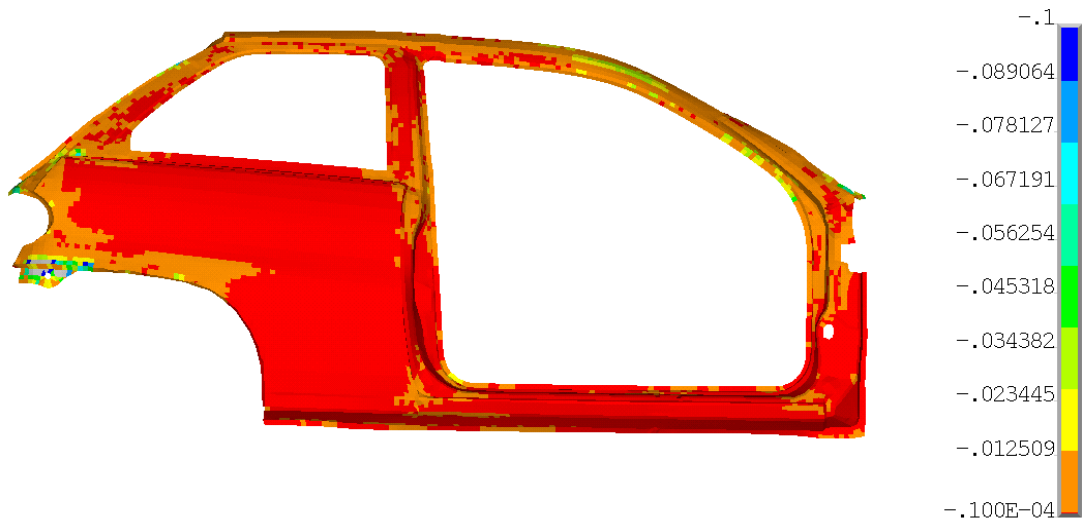

(b) The results of calculating modal flexibility sensitivity using PM

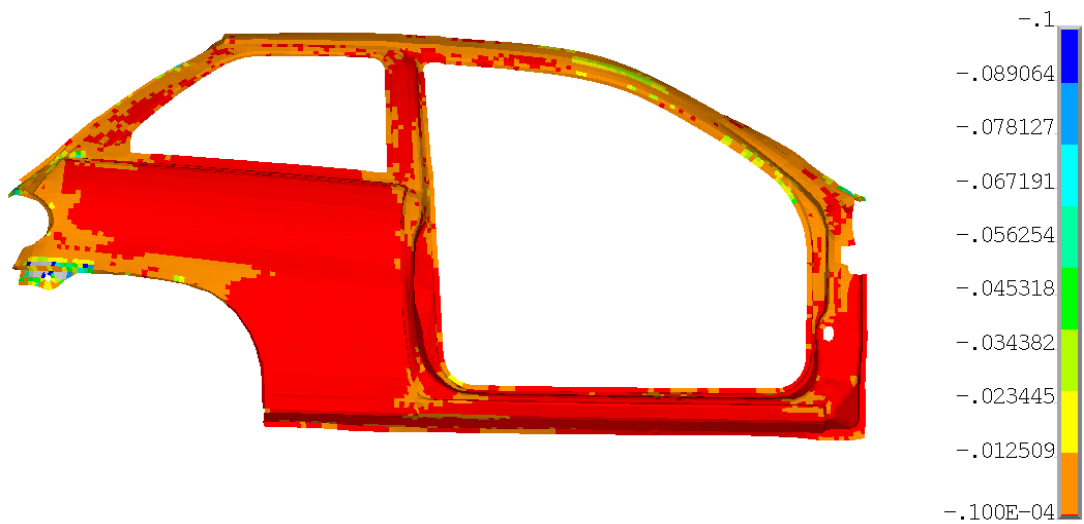

(c) The results of calculating modal flexibility sensitivity using PM

**FIGURE 8** The sensitivity results for the Side Outer Panel is calculated using three different methods.

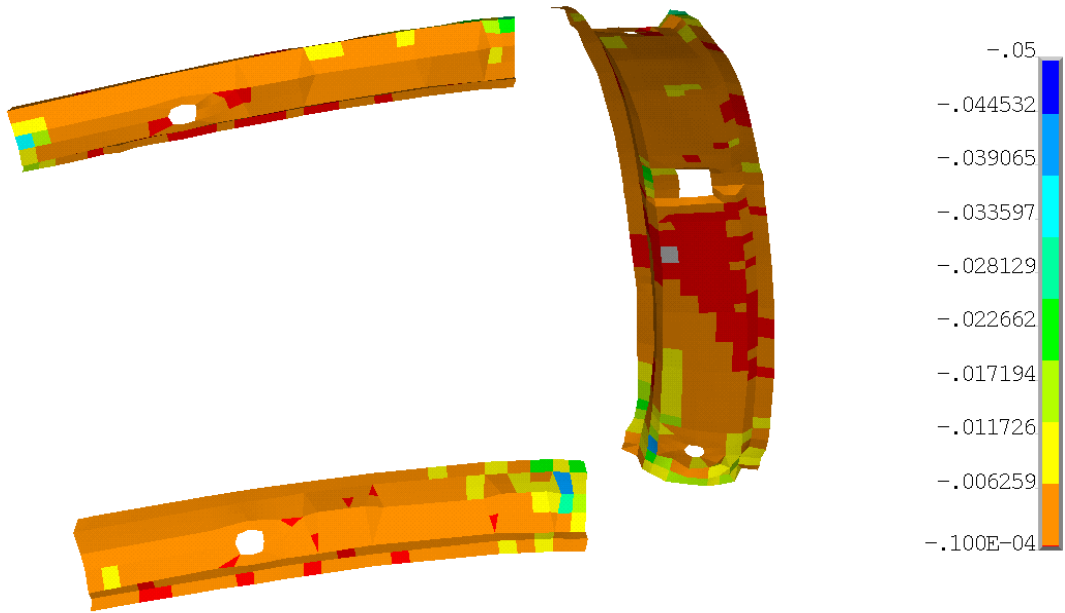

(a) The results of calculating modal flexibility sensitivity using PM

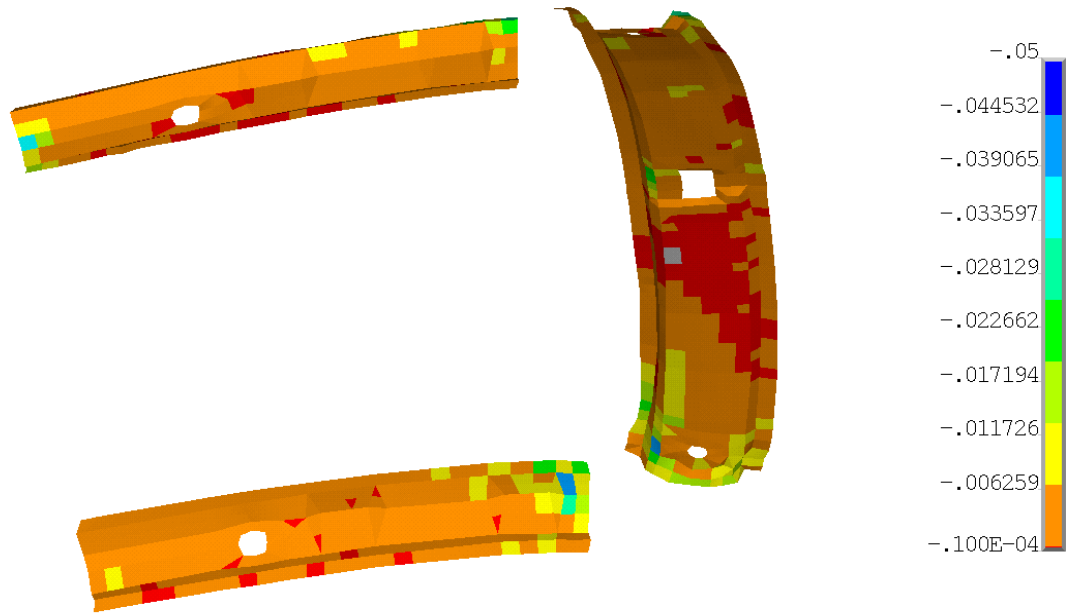

(b) The results of calculating modal flexibility sensitivity using ADAM

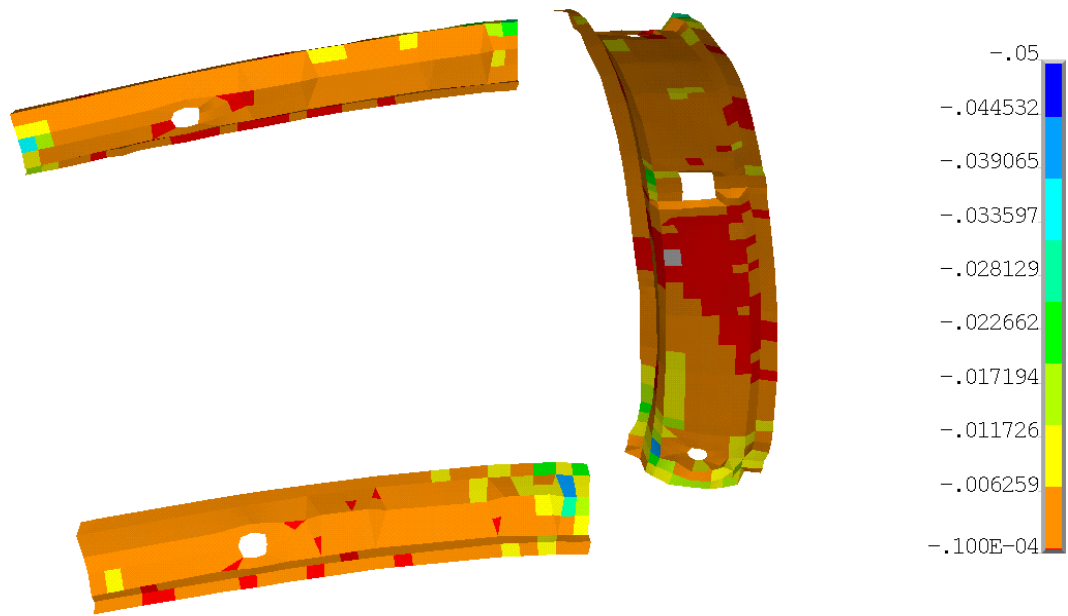

(c) The results of calculating modal flexibility sensitivity using ADNE

**FIGURE 9** The sensitivity results for Reinforcement Bracket is calculated using three different methods.

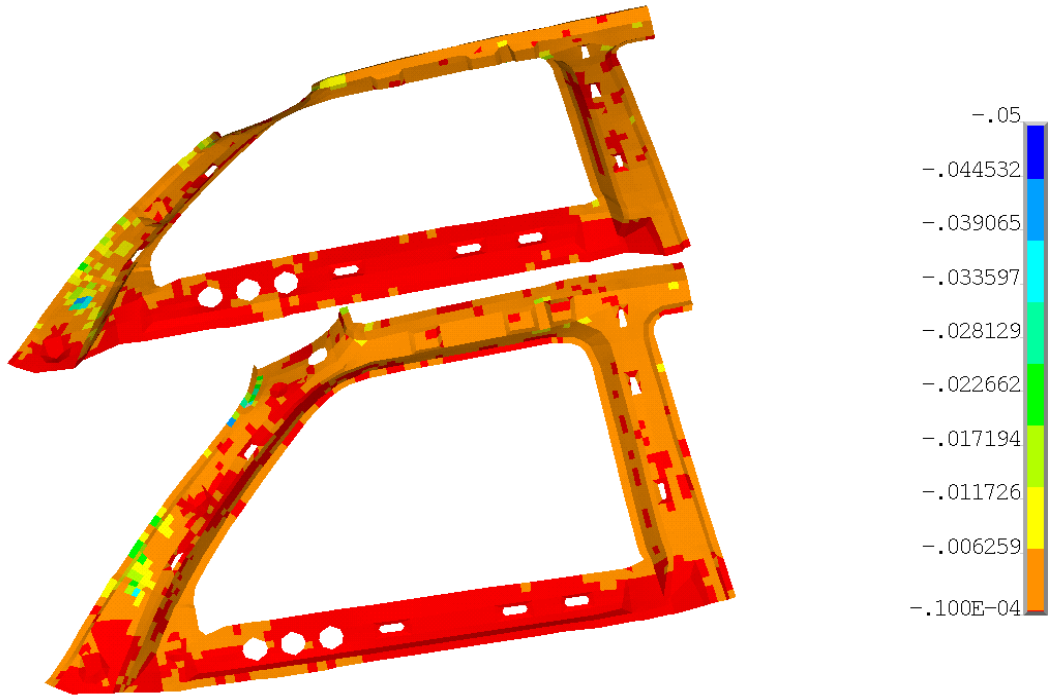

(a) The results of calculating modal flexibility sensitivity using PM

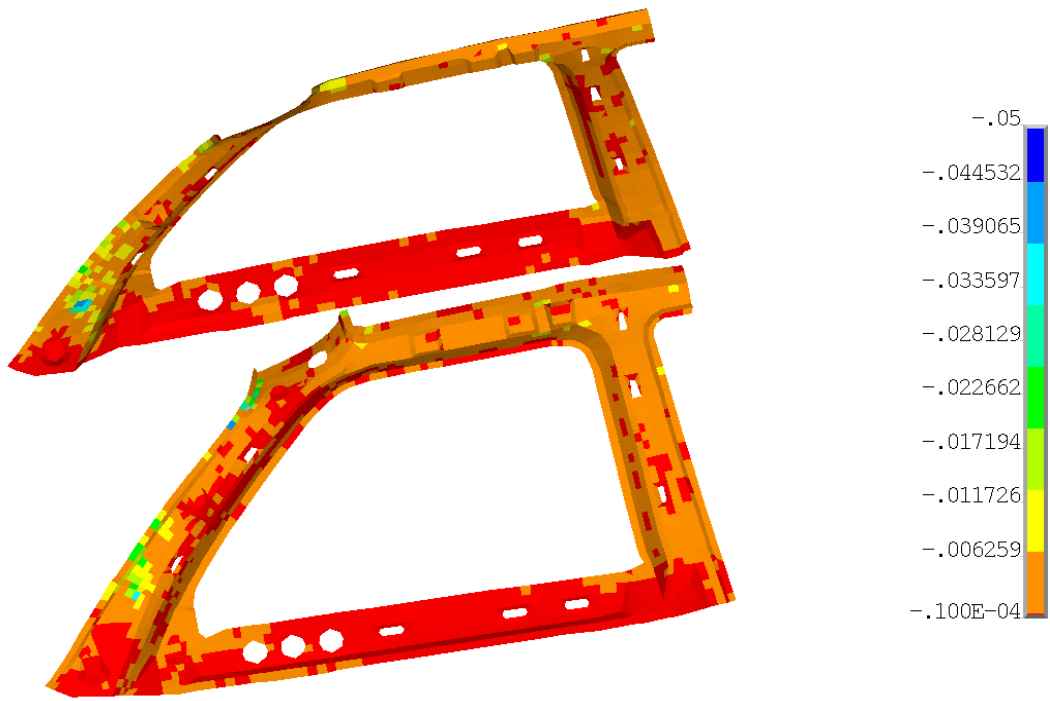

(b) The results of calculating modal flexibility sensitivity using ADAM

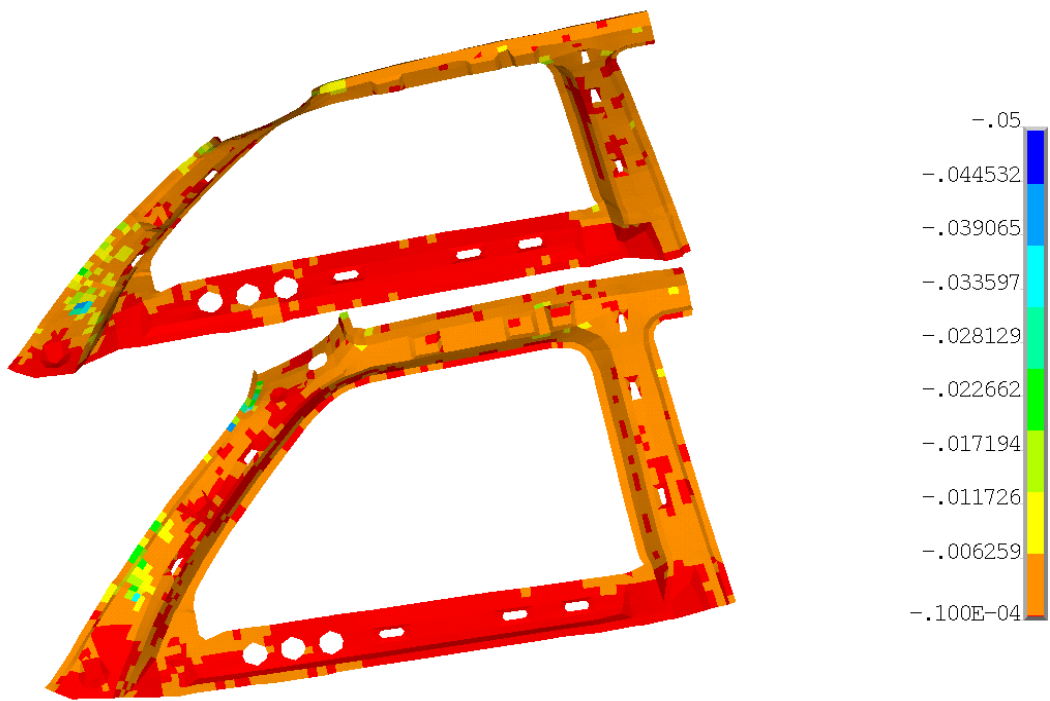

(c) The results of calculating modal flexibility sensitivity using ADNE

**FIGURE 10** The sensitivity results for Window Frame is calculated using three different methods.

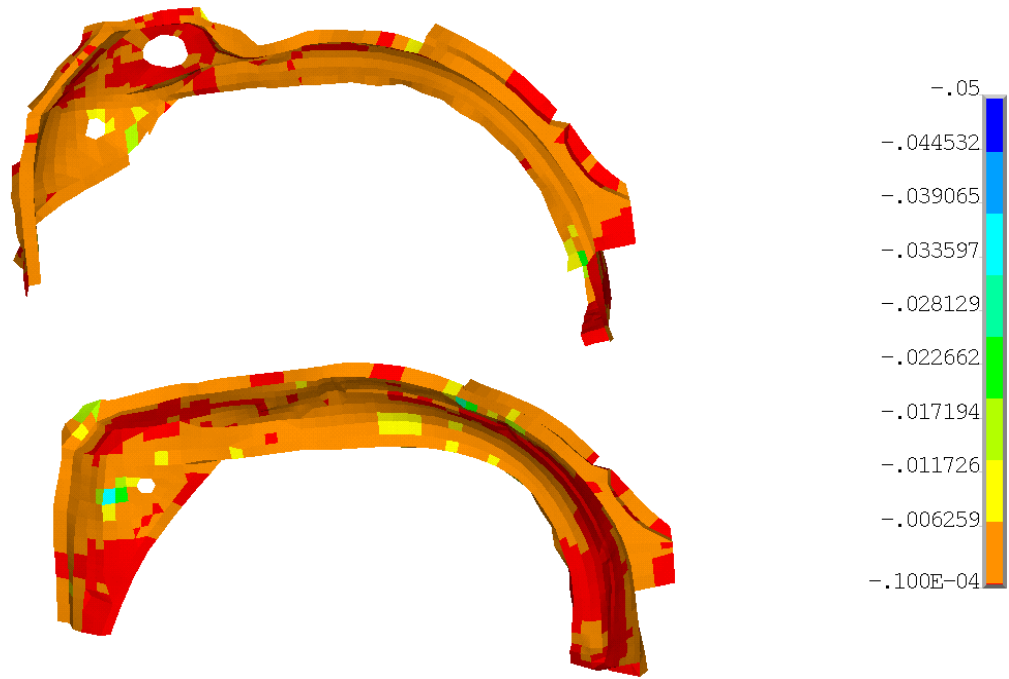

(a) The results of calculating modal flexibility sensitivity using PM

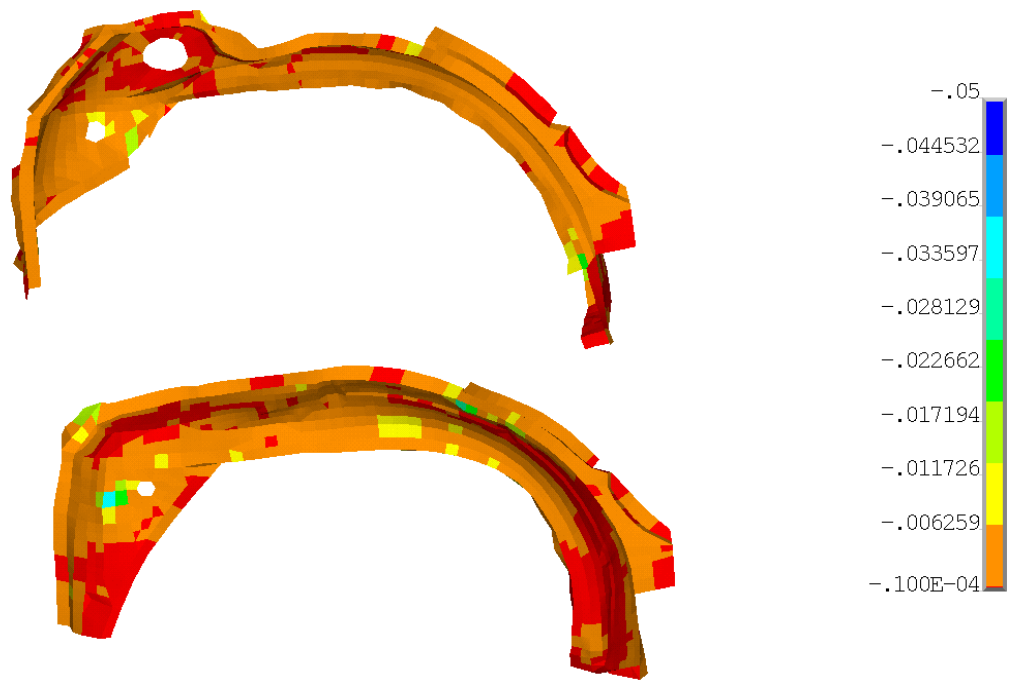

(b) The results of calculating modal flexibility sensitivity using ADAM

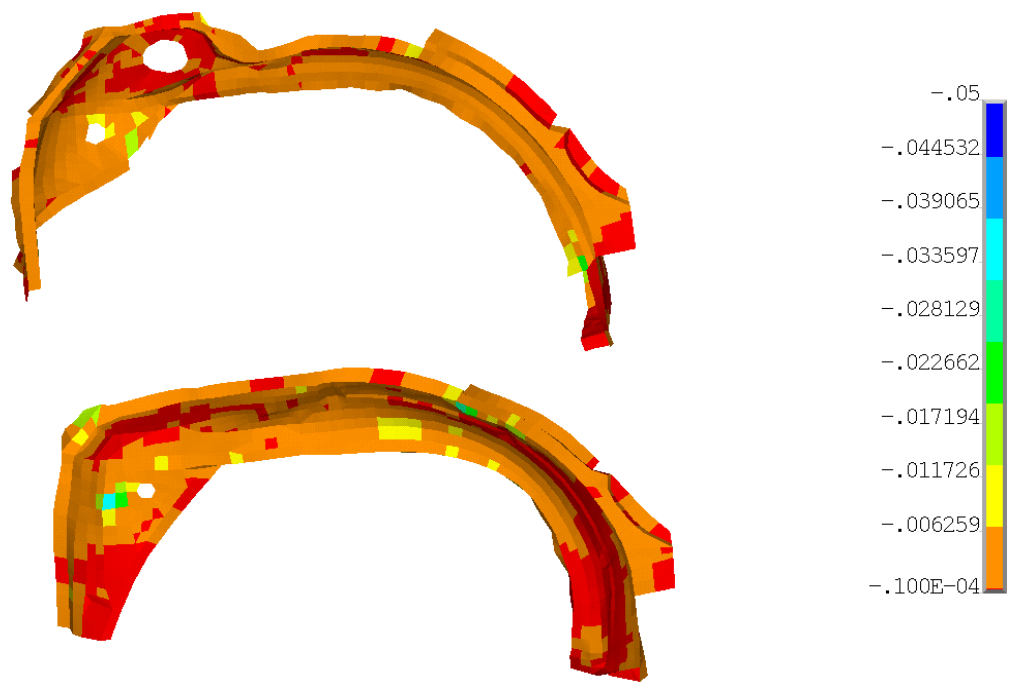

(c) The results of calculating modal flexibility sensitivity using ADNE

**FIGURE 11** The sensitivity results for the Fender Liner and is calculated using three different methods.

6. **Conclusions**

In this paper, an efficient framework is proposed for solving the sensitivity of structural dynamic characteristics related to eigenmode with respect to multiple variables. Firstly, a formulation of a non-singular coefficient matrix and an algebraic method for calculating the derivatives of eigenvectors is developed to simplify the expression for sensitivity calculations. Subsequently, based on this new expression for eigenmode sensitivity, a framework for sensitivity analysis of structural dynamic characteristics related to eigenmodes with multiple parameters is established. With the incorporation of a preconditioning iterative method, the new computational framework effectively enhances the computational efficiency. This framework retains the advantages of the adjoint vector method while avoiding fill-in operations, which is crucial for large-scale computations, thereby significantly reduce CPU computational time. It can be integrated into a structural optimization and other CAE software. Further extensions of the present new framework to address the problems of damped systems will be considered.

**Acknowledgements**

This work was supported in part by China Natural National Science Foundation (grant No. 22341302), the National Key Research and Development Program of China (grant No. 2020YFA0713602, 2023YFA1008803), and the Key Laboratory of Symbolic Computation and Knowledge Engineering of Ministry of Education of China housed at Jilin University.

**Declaration of Competing Interest**

The authors declare that they have no known competing financial interests or personal relationships that could have appeared to influence the work reported in this paper.

**CRediT authorship contribution statement**

**Kai huang :** Writing – original draft, Program and Calculation. **Zhengguang Li:**

Writing – review &editing, Methodology, Conceptualization. **Xiuli Wang**: Writing – review & editing, Methodology.

**Appendix A**

Denote the eigenvector matrix of generalized eigenvalue problems Eq. (1) is :

$$\mathbf{\Phi} = [\boldsymbol{\varphi}_1, \boldsymbol{\varphi}_2, \cdots, \boldsymbol{\varphi}_i, \cdots, \boldsymbol{\varphi}_n], \tag{A.1}$$

Premultiplying $\mathbf{G}$ by $\mathbf{\Phi}^{\mathrm{T}}$ and postmultiplying $\mathbf{\Phi}$ we obtain

$$\mathbf{\Phi}^{\mathrm{T}}\mathbf{G}\mathbf{\Phi} = \mathbf{\Phi}^{\mathrm{T}}(\mathbf{K} - \lambda_i \mathbf{M} + \mathbf{M}\boldsymbol{\varphi}_i \boldsymbol{\varphi}_i^T \mathbf{M})\mathbf{\Phi}, \tag{A.2}$$

Using Eq. (1), Eq. (2) we have

$$\mathbf{\Phi}^{\mathrm{T}}\mathbf{G}\mathbf{\Phi} = \mathbf{\Phi}^{\mathrm{T}}(\mathbf{K} - \lambda_i \mathbf{M})\mathbf{\Phi} + \mathbf{\Phi}^{\mathrm{T}}\mathbf{M}\boldsymbol{\varphi}_i \boldsymbol{\varphi}_i^T \mathbf{M}\mathbf{\Phi}$$
$$= \begin{bmatrix} \Lambda_I & 0 & 0 \\ 0 & 0 & 0 \\ 0 & 0 & \Lambda_h \end{bmatrix} + \begin{bmatrix} 0 & 0 & 0 \\ 0 & 1 & 0 \\ 0 & 0 & 0 \end{bmatrix}, \tag{A.3}$$

where $\Lambda_I = diag(\lambda_1 - \lambda_i, \ldots, \lambda_{i-1} - \lambda_i)$, $\Lambda_h = diag(\lambda_{i+1} - \lambda_i, \ldots, \lambda_n - \lambda_i)$, obviously $\mathbf{G}$ is invertible when the eigenvalues $\lambda_i$ is distinct.